\documentclass[twocolumn,english,superscriptaddress,floatfix,longbibliography]{revtex4-1}
\usepackage[T1]{fontenc}
\usepackage[latin9]{inputenc}
\usepackage{xcolor}
\usepackage{pdfcolmk}
\usepackage{amsmath}
\usepackage{amssymb}
\usepackage{graphicx}
\PassOptionsToPackage{normalem}{ulem}
\usepackage{ulem}

\makeatletter

\providecolor{lyxadded}{rgb}{0,0,1}
\providecolor{lyxdeleted}{rgb}{1,0,0}
\DeclareRobustCommand{\lyxsout}[1]{\ifx\\#1\else\sout{#1}\fi}

\usepackage{upgreek}

\makeatother

\usepackage{babel}
\begin{document}
\title{Inducing exceptional points, enhancing plasmon quality and creating
correlated plasmon states with modulated Floquet parametric driving}
\author{Egor I. Kiselev}
\affiliation{Physics Department, Technion, 320003 Haifa, Israel}
\affiliation{The Helen Diller Quantum Center, Technion, Haifa 3200003, Israel}
\author{Mark S. Rudner}
\affiliation{Department of Physics, University of Washington, Seattle, Washington
98195-1560, USA}
\author{Netanel H. Lindner}
\affiliation{Physics Department, Technion, 320003 Haifa, Israel}
\begin{abstract}
We propose a method to parametrically excite low frequency collective
modes in an interacting many body system using a Floquet driving at
optical frequencies with a modulated amplitude. We demonstrate that
it can be used to design plasmonic time-varying media with singular
dispersions. Plasmons near resonance with half the modulation frequency
exhibit two lines of exceptional points connected by dispersionless
states. Above a critical driving strength, resonant plasmon modes
become unstable and undergo a continuous transition towards a crystal-like
structure stabilized by interactions and nonlinearities. This new
state breaks the discrete time translational symmetry of the drive
as well as the translational and rotational spatial symmetries of
the system and exhibits soft, Goldstone-like phononic excitations.
Below the instability threshold, our method can be used to enhance
the quality of plasmon resonances.
\end{abstract}
\maketitle

\section{Introduction}

The collective modes of a many-body system fingerprint the symmetries
underlying a phase of matter. They also play an important role in
transitions to non-equilibrium states when the system is subjected
to an external drive. Resonantly driving collective modes can lead
to exotic wave propagation effects and, at strong driving, to instabilities
leading to new symmetry breaking non-equilibrium states \citep{cavalleri2018photo_induced_superconductivity,fausti2011light_induced_superconductivity,forst2011driving_magnetic_order_lattice_excitation,rini2007control_electr_phase_vibrational_excitation,rudner2019_berryogenesis}.
An example for such an instability in classical physics is Faraday
waves formed by parametrically driven surface waves \citep{faraday1831_patterns,benjamin1954_faraday_stability,kumar1994_faraday}.

Accessing such phenomena in electronic systems is challenging due
to a lack of methods for the control and manipulation of collective
modes in solids. At the same time, realizing such control holds a
lot of promise for applications in fields like plasmonics or spintronics.
Here, we show that Floquet engineering, which employs driving at optical
frequencies to manipulate electronic bandstructures \citep{oka2009photovoltaic_hall_effect,kitagawa2011floquetinduced,lindner2011floquet,wang2013_Floquet-Bloch_states_observation,mahmood2016selective_scattering_floquet-bloch_volkov,mciver2020light_anomaouls_Hall_graphene,kitagawa2010topological_characterization_driven_quantum_system,zhou2023black_phosphorus_floquet,usaj2014_floquet_graphene_topo,perez2014floquet_traphene_topo,oka2019floquet_review,katz2020optically,esin2018q_steady_state_topo_ins,esin2020floquet_metal_insulator,esin2021_liquid_crystal,dehghani2015_floquet_topo,genske2015floquet_boltzmann,glazman1983kinetics_pulses_semiconductor,dehghani2014dissipative_topo_floquet,sentef2015pump_probe_floquet,chan2016floquet_ref,farrell2015floquet_ref,gu2011floquet_ref,hubener2017floquet_ref,jiang2011floquet_ref,kennes2019floquet_ref,kundu2013floquet_ref,thakurathi2017floquet_ref,frank2013_lattice_modulations_metal_insulator,castro2022optimal_floquet_control},
can be an effective tool to control collective modes. We focus on
plasmons in clean, two dimensional materials which exhibit large quality
factors due to the absence of impurity scattering and the suppression
of Landau damping by the $\sqrt{q}$ shape of their dispersion. We
demonstrate that optical driving fields, far off-resonant relative
to the plasmon frequencies, can lead to the formation of exceptional
points and non-dispersive plasmonic states through a parametric coupling
mechanism which we call \textquotedblleft modulated Floquet parametric
driving'' (MFPD). For strong driving, we predict the existence of
an emergent non-equilibrium state with spatio-temporal correlations
that break time and space translation symmetry. This connects to the
ongoing interest in using complex multi-component driving signals
to control solid-state electrons \citep{neufeld2019high_harmonic_multiple_timescale,neufeld2019floquet_polarization,castro2022optimal_floquet_control,ikeda2022floquet_two_component}
and to induce non-equilibrium correlated states \citep{basov2017towards_on_demand,bloch2022strongly_correlated_light_induced,mentink2015_mott_floquet_exchange,rudner2019_berryogenesis,esin2021_liquid_crystal,dykman2018_time_symmetry_breaking,kim2006_old_time_crystal1,heo2010_old_time_crystal2,kyprianidis2021observation_discrete_time_crystal}.
Furthermore we show that MFPD can be used to increase plasmon quality
factors.

\begin{figure}[t]
\centering{}\includegraphics[width=0.95\columnwidth]{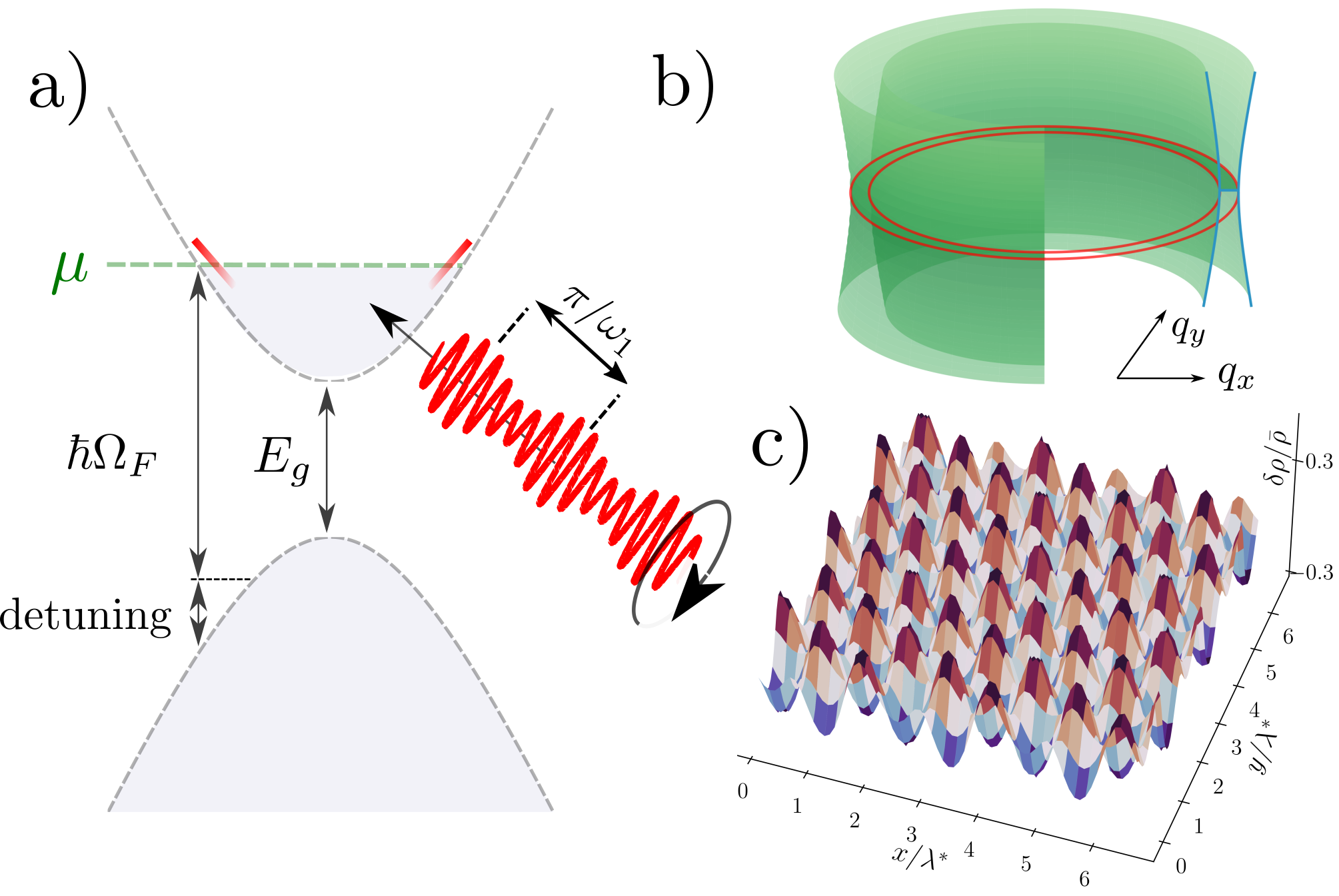}\caption{Modulated Floquet parametric driving, exceptional lines and non-equilibrium
steady states: a) Illustration of the quasi-energy bands of electrons
in a gapped system under MFPD. The dashed gray lines represent the
undriven band structure. A Floquet signal with the frequency $\Omega_{F}$,
where $\hbar\Omega_{F}>E_{g}$, is applied. The quasi-energy bands
in the vicinity of the Fermi surface for a non-zero driving amplitude
are shown in red. By modulating the amplitude of the fast Floquet
drive with a frequency $2\omega_{1}$, an oscillation of the dispersion
and the effective mass of the electrons at the Fermi level is induced
(see Eq. (\ref{eq:Oscillating_mass})). b) The periodic modulation
of the Fermi velocity induces two lines of exceptional points (shown
as red circles) in the rotationally symmetric dispersion of the soft
THz plasmon mode (green surfaces) -- see Fig. \ref{fig:Onset-of-the_instability}
for an in-detail cross-sectional view. c) Above a critical value,
the mass oscillation parametrically excites THz plasmons. Due to nonlinear
and interaction effects, the plasmons arrange themselves in a crystal-like
structure whose periodicity is determined by the condition $\omega_{1}=\omega_{\mathrm{pl}}\left(q^{*}\right)$,
where $\omega_{\mathrm{pl}}\left(q\right)$ is the plasmon dispersion.
The crystalline state breaks the rotational and translational symmetries
of the undriven system and supports soft phonon-like Goldstone modes.
\label{fig:dispersions_and_setup}}
\end{figure}

MFPD takes advantage of the fact that the electronic properties of
two dimensional materials are modified in the presence of a high frequency
time-periodic driving fields \citep{rudner2020band_engineering}.
Slowly modulating the parameters of these fields, an electronic system
can be controlled dynamically. As illustrated in Fig.\ref{fig:dispersions_and_setup}a),
a high-frequency Floquet driving alters the shape of the quasi-energy
bands (red curves). In particular, the effective mass $m^{*}$ of
electrons at the Fermi surface changes. This change results in an
altered plasmon dispersion. A periodic modulation of the Floquet drive's
amplitude results in a periodically changing mass that parametrically
excites plasmon modes. This effect induces two lines of exceptional
points in the plasmon dispersion {[}Fig.\ref{fig:dispersions_and_setup}b){]}.
These exceptional lines are separated by a non-dispersive gap where
the real part of the plasmon dispersion is flat. Inside the gap, damping,
which is reflected in the negative imaginary part of the plasmon dispersion,
is reduced by MFPD. This effect can be used to enhance the quality
of the plasmon resonance and increase plasmonic propagation lengths.

For sufficiently strong driving, for plasmons with wavenumbers $q^{*}$
meeting the condition $\mathrm{Re}\left[\omega_{\mathrm{pl}}\left(q^{*}\right)\right]=\omega_{1}$,
where $\omega_{\mathrm{pl}}\left(q\right)$ is the plasmon dispersion
and $2\omega_{1}$ is the amplitude modulation frequency, the imaginary
part of the dispersion changes sign. These plasmons become unstable
and, after a short period of exponential growth, form a crystal-like
lattice with a periodically modulated electron density {[}Fig.\ref{fig:dispersions_and_setup}
c){]}. The structure of this crystal is determined by the nonlinearities
of the system. The density oscillates at half the modulation frequency,
thus breaking the $\pi/\omega_{1}$ discrete time translation symmetry
of the drive -- a behavior known in discrete time crystals \citep{kim2006_old_time_crystal1,heo2010_old_time_crystal2,else2016floquet_discrete_time_crystals,yao2017discrete_time_crystal_original,kyprianidis2021observation_discrete_time_crystal,zhang2017observation_discrete_time_crystal,choi2017observation_discrete_time_crystal,natsheh2021critical_properties_time_crystal,dykman2018_time_symmetry_breaking}.

An important challenge in Floquet engineered systems is heating \citep{lazarides2014_floquet_heating,dalessio2014_floquet_heating,bukov2016heating_floquet,bukov2015_floquet_heating_prethermal,else2017_floquet_heating_prethermal,mori2018_floquet_heating_prethermal,reitter2017_floquet_heating_experiment,singh2019_floquet_heating_experiment,galitsky1970_driven_semiconductor_steady_state,shirai2015condition_for_gibbs_in_floquet_steady_state,seetharam2015baths_controlled_floquet_population,iadecola2015occupation_topo_floquet,seetharam2019floquet_steady_state_interacting,liu2015classification_of_floquet_statistical_distribution,abanin2015exponentially_slow_heating_floquet}.
Here, we show that the parametric excitation of plasmons by MFPD can
be achieved in an off-resonant regime where (momentum conserving)
single-photon excitation of electrons is blocked. To achieve this,
the base-tone drive frequency $\Omega_{F}$ and the Fermi energy $\mu$
must be chosen such that all electronic states supporting a resonant
interband single photon excitation with an energy transfer of $\hbar\Omega_{F}$
and near-zero momentum transfer lie below the Fermi surface (see shaded
regions Fig. \ref{fig:dispersions_and_setup}). Processes involving
multiple photons are suppressed in the small ratio of drive amplitude
and drive frequency, leading to strongly reduced heating rates due
to photoexcited electrons. We estimate the heating due to momentum
non-conserving absorption mediated by phonons or impurity-scattering,
as well as interaction assisted processes, and find that the heating
rates are small, such that they can be balanced by the cooling power
of the crystal lattice.

Below, we describe in more detail the creation of singular plasmon
dispersions and the excitation of plasmonic instabilities by MFPD,
the transition into the crystalline state and the propagation of Goldstone-like,
phononic collective modes in the symmetry broken state.

\section{Results}

\subsection{Modulated Floquet parametric driving}

The idea of MFPD is to use the dependence of the quasi-energy band
structure on the parameters of the drive to couple to the soft collective
modes of the system. We show how the slow modulation of the driving
amplitude leads to a time-varying effective electron mass. For concreteness,
let us consider a coherently driven, gapped 2D Dirac system described
by the Hamiltonian
\begin{equation}
H=\sum_{\mathbf{k}}\mathbf{c}_{\mathbf{k}}^{\dagger}\left[H_{0}\left(\mathbf{k}\right)+H_{d}\left(t\right)\right]\mathbf{c}_{\mathbf{k}}+\sum_{\mathbf{q}}V\left(\mathbf{q}\right)\hat{\rho}_{\mathbf{q}}\hat{\rho}_{-\mathbf{q}}.\label{eq:general_H}
\end{equation}
We will first illustrate how the external drive modifies the dispersion
in a single valley. The dispersion in the opposite valley is modified
similarly. Both valleys will be taken into account when considering
the collective modes. For small values of k around the center of the
valley, we have $H_{0}=\mathbf{d}\cdot\boldsymbol{\sigma}$, where
$\mathbf{d}=\left[\lambda k_{x},\lambda k_{y},E_{g}/2\right]$, $\boldsymbol{\sigma}$
is a Pauli matrix vector describing the orbital pseudospin degree
of freedom and $E_{g}$ is the energy gap between the two bands. $\mathbf{c}_{\mathbf{k}}^{\dagger}$
($\mathbf{c}_{\mathbf{k}}$) are two-component electron creation (annihilation)
operators, where individual components describe electrons with different
pseudospin orientations, and $H_{d}\left(t\right)=e\mathbf{A}\left(t\right)\cdot\nabla_{\mathbf{k}}H_{0}$
is the driving Hamiltonian derived from minimal coupling.

We assume circularly polarized light with an amplitude $\mathcal{E}$
described by $\mathbf{A}\left(t\right)=\left(\mathcal{E}/\Omega_{F}\right)\left[-\sin\Omega_{F}t,\cos\Omega_{F}t,0\right]$.
The 2D Fourier transform of the Coulomb potential is given by $V\left(\mathbf{q}\right)=2\pi/q$
and $\hat{\rho}_{\mathbf{q}}=\sum_{\mathbf{k}}\mathbf{c}_{\mathbf{k}}^{\dagger}\mathbf{c}_{\mathbf{k}+\mathbf{q}}$
is the density operator. We emphasize that we use the gapped Dirac
Hamiltonian of Eq. (\ref{eq:general_H}) as an example. The physics
presented here does not depend on the precise band structure of the
system.

It is convenient to work in a rotating frame defined by the unitary
transformation $U\left(t\right)=e^{i\hat{\mathbf{d}}\cdot\boldsymbol{\sigma}\Omega_{F}t/2}$,
where the spectrum of the transformed single-particle part of Eq.
(\ref{eq:general_H}) is given by 
\begin{equation}
\varepsilon_{k}\approx\sqrt{\left(\left|\mathbf{d}\right|-\hbar\frac{\Omega_{F}}{2}\right)^{2}+\frac{e^{2}\mathcal{E}^{2}\lambda^{2}}{4\Omega_{F}^{2}\hbar^{2}}\left(2-\frac{E_{g}}{\left|\mathbf{d}\right|}\right)}\label{eq:rot_frame_dispersion-1}
\end{equation}
for $\lambda k/\left|\mathbf{d}\right|\ll1$. In writing Eq. (\ref{eq:rot_frame_dispersion-1})
we neglected time dependent terms in the rotating frame \citep{lindner2011floquet}.
In this frame and with the approximations made, the time dependence
caused by the fast $\Omega_{F}$ oscillation of the Floquet drive
does not explicitly appear in the equations.

We consider the metallic regime with the Fermi surface lying in the
upper band (see Fig. \ref{fig:dispersions_and_setup}) and expand
the spectrum around $k_{F}$ -- the Fermi momentum of the undriven
system. We write:
\begin{equation}
\varepsilon_{k}\approx\hbar v_{F}\left(\mathcal{E},\Omega_{F}\right)\left(k-k_{F}\right)+\varepsilon_{k_{F}}\left(\mathcal{E},\Omega_{F}\right)\label{eq:rot_frame_dispersion}
\end{equation}
The Fermi velocity $v_{F}\left(\mathcal{E},\Omega_{F}\right)$ depends
on the amplitude and frequency of the Floquet drive.

A variation of the drive amplitude $\mathcal{E}$ results in a small
change of the dispersion $\varepsilon_{k}\rightarrow\bar{\varepsilon}_{k}+\delta\varepsilon_{k}$
near the Fermi surface. To clearly distinguish between constant quantities
and quantities oscillating with the slow modulation frequency $\omega_{1}$,
here and in the following, we write the constant part with a bar.
The total charge of the system is conserved and therefore $k_{F}$
is fixed. However, the slope of $\varepsilon_{k}$ at $k_{F}$ and
therefore the effective mass of the electrons $m^{*}=\hbar k_{F}/v_{F}\left(\mathcal{E},\Omega_{F}\right)$
are altered. For a small $\delta\varepsilon_{k}$, we find 
\begin{equation}
\bar{\varepsilon}_{k}+\delta\varepsilon_{k}\approx\left(1-\frac{\delta m^{*}}{\bar{m}^{*}}\right)\frac{\hbar^{2}k_{F}}{\bar{m}^{*}}\left(k-k_{F}\right).
\end{equation}

We consider a slow, adiabatic oscillation of the electric field amplitude
$\mathcal{E}$, 
\begin{equation}
\mathcal{E}\left(t\right)=\bar{\mathcal{E}}+\delta\mathcal{E}\cos\left(2\omega_{1}t\right),\label{eq:small_change_of_V_time}
\end{equation}
such that $\omega_{1}\ll\Omega_{F}$. This slow oscillation does not
couple to any single-electron degrees of freedom. However, as demonstrated
below, it does couple to the soft plasmon mode through a parametric
resonance induced by the periodic change of the effective mass
\begin{equation}
m^{*}\left(t\right)=\bar{m}^{*}\left(1+\frac{\delta m^{*}\left(t\right)}{\bar{m}^{*}}\right).\label{eq:Oscillating_mass}
\end{equation}

\subsection{Equations of motion for collective dynamics\label{subsec:Equations-of-motion}}

To describe the electron dynamics of MFPD driven plasmons, we use
a hydrodynamic description of Coulomb interacting electrons (see,
e.g., Refs. \citep{Eguiluz1976hydrodynamicPlasmons,PinesNozieres1,Briskot2015}).
The equations of motion follow from the conservation of charge and
momentum \citep{Forster,LLHydro,lucas2015memory}. The continuity
equation for the electron density $\rho$ is
\begin{equation}
\partial_{t}\rho\left(t,\mathbf{x}\right)=-\partial_{i}\left[\rho\left(t,\mathbf{x}\right)u_{i}\left(t,\mathbf{x}\right)\right],\label{eq:conti}
\end{equation}
where $\mathbf{u}\left(t,\mathbf{x}\right)$ is the electron flow
velocity. The continuity equation for the momentum density is given
by the Euler equation
\begin{equation}
\partial_{t}\left(m^{*}\rho u_{i}\right)+\partial_{j}\Pi_{ij}=-\gamma m^{*}\rho u_{i}-e\rho\partial_{i}\phi,\label{eq:nav_stokes}
\end{equation}
where $m^{*}=\hbar k_{F}/v_{F}$ is the effective mass of electrons
at the Fermi surface, given by the ratio of Fermi momentum $\hbar k_{F}$
and Fermi velocity $v_{F}$, $u_{i}$ is the electron flow velocity,
and $\gamma$ is the rate of momentum relaxation, e.g., due to impurity
or phonon scattering. The stress tensor $\Pi_{ij}$ is given by 
\begin{equation}
\Pi_{ij}=m^{*}\rho u_{i}u_{j}+\delta_{ij}p,\label{eq:stress_tensor}
\end{equation}
where $p$ is the pressure. The electrostatic potential is given by
\begin{equation}
\phi\left(t,\mathbf{x}\right)=\int d^{2}x'\frac{e\rho\left(t,\mathbf{x}'\right)}{4\pi\varepsilon\left|\mathbf{x}-\mathbf{x}'\right|}.\label{eq:poisson_eq}
\end{equation}
Here $e$ is the electron charge and $\varepsilon$ the effective
dielectric constant. Solutions of equations (\ref{eq:conti}) and
(\ref{eq:nav_stokes}) describe the collective oscillations of the
electron fluid.

\subsection{Oscillating mass\label{subsec:Oscillating-mass}}

The oscillating mass increment $\delta m^{*}\left(t\right)$ in Eq.
(\ref{eq:nav_stokes}) acts as a parametric drive, which is known
to lead to instabilities \citep{Landau_Lifshitz_Mechanics}. In the
following we identify the parametric instability of the charge density
$\rho$. It is convenient to take the divergence of Eq. (\ref{eq:nav_stokes})
and to combine the result with the continuity equation (\ref{eq:conti}),
including the drive-induced temporal modulation of $m^{*}\left(t\right)$.
Doing so, we find
\begin{equation}
\partial_{t}m^{*}\left(t\right)\partial_{t}\rho+\gamma m^{*}\left(t\right)\partial_{t}\rho-m^{*}\left(t\right)\partial_{i}\partial_{j}\rho u_{i}u_{j}=\partial_{i}\rho\partial_{i}\phi,\label{eq:rho_nonlin_eq}
\end{equation}
where we neglected the pressure term $p$ since its contribution is
subleading to the long-range Coulomb potential \citep{Lucas20182D,kiselev2021_superdiffusive_modes}\footnote{The Fourier transform of the Coulomb potential is $V\left(q\right)=2\pi/q$
and is responsible for the characteristic $\sim\sqrt{q}$ dispersion
of two dimensional plasmons, whereas the pressure term scales as $\sim q$
and will only contribute a small correction.}. Unless otherwise indicated, throughout the paper, derivatives act
on all functions to the right. In our derivations, we also neglect
the time dependent contribution to the damping term, which is negligible
in comparison to the static one.

\subsection{Instabilities and exceptional points\label{subsec:Instabilities-and-exceptional}}

In this section, we present how the oscillating mass under MFPD in
Eq.(\ref{eq:Oscillating_mass}) influences the plasmon dynamics under
Eq. (\ref{eq:rho_nonlin_eq}). Consider the electron density $\rho=\bar{\rho}+\delta\rho$,
where $\bar{\rho}$ is the equilibrium electron density and $\delta\rho$
is a perturbation of $\bar{\rho}$. To linear order in $\delta\rho$,
and performing a Fourier transform $\mathbf{x}\rightarrow\mathbf{q}$,
for a modulation $\delta m^{*}\left(t\right)/\bar{m}^{*}=h\cos\left(2\omega_{1}t\right)$,
we obtain from Eq. (\ref{eq:rho_nonlin_eq})
\begin{equation}
\partial_{t}\left(1+h\cos\left(2\omega_{1}t\right)\right)\partial_{t}\delta\rho_{q}+\gamma\partial_{t}\delta\rho_{q}+\omega_{\mathrm{pl}}^{2}\left(q\right)\delta\rho_{q}=0.\label{eq:param_plasmon_driving}
\end{equation}
The plasmon dispersion is given by 
\begin{equation}
\omega_{\mathrm{pl}}\left(q\right)=\sqrt{\frac{e\bar{\rho}}{4\pi\bar{m}^{*}\varepsilon}q^{2}V\left(q\right)}.\label{eq:plasmon_dispersion}
\end{equation}
The parameter $\gamma$ captures scattering on impurities and lattice
excitations, which lead to damping of the plasmon modes. Intrinsic
Landau damping is strongly suppressed for the wavenumbers and frequencies
under consideration, since they lie outside of the particle-hole continuum.
Indeed, recent experiments on high quality graphene showed that damping
mainly arises due to scattering with phonons \citep{ni2018_plasmon_quality_factor_graphene}.
It is important to note that this scattering mechanism is strongly
suppressed at low temperatures, where dissipation is dominated by
losses caused by electric fields penetrating into the substrate \citep{ni2018_plasmon_quality_factor_graphene}.
In general, it is believed that substrate engineering and strong cooling
could strongly increase the plasmon quality \citep{principi2013intrinsic_graphene_plasmon_quality_factor}.
Notice that Eq. (\ref{eq:param_plasmon_driving}) is equivalent to
the equation describing the propagation of the magnetic field in photonic
time crystals \citep{lyubarov2022photonic_time_crystal_amplified}.

In Sec. \ref{subsec:Slowly-varying-envelope}, we derive a solution
for $\delta\rho_{q}$ using the slowly varying envelope approximation,
which is valid near the parametric resonance where $\omega_{\mathrm{pl}}^{2}\left(q\right)\approx\omega_{1}$:
\begin{equation}
\delta\rho_{q}\left(t\right)=a_{q}\left(t\right)\cos\left(\omega_{1}t\right)+b_{q}\left(t\right)\sin\left(\omega_{1}t\right).\label{eq:line_delta_rho_sol}
\end{equation}
Here the amplitudes are given by $a_{q}\left(t\right)=a_{q}\left(0\right)e^{s_{\pm}\left(q\right)t}$
and $b_{q}\left(t\right)=b_{q}\left(0\right)e^{s_{\pm}\left(q\right)t}$,
where
\begin{equation}
s_{\pm}\left(q\right)=-\frac{\gamma}{2}\mp i\frac{\omega_{1}}{2}\sqrt{\frac{\left[\omega_{\mathrm{pl}}^{2}\left(q\right)-\omega_{1}^{2}\right]^{2}}{\omega_{1}^{4}}-\left(\frac{h}{2}\right)^{2}}.\label{eq:s}
\end{equation}

We now examine the plasmon dispersion in the presence of parametric
driving. To this end, it is convenient to rewrite Eq. (\ref{eq:line_delta_rho_sol}):
according to Floquet's theorem, any solution to Eq. (\ref{eq:param_plasmon_driving})
can be written in the form $\delta\rho_{q}=e^{-i\Lambda\left(q\right)t}u_{\Lambda}\left(t\right)$,
where $u_{\Lambda}\left(t+\pi/\omega_{1}\right)=u_{\Lambda}\left(t\right)$.
Choosing $u_{\Lambda}\left(t\right)=\left[a_{\Lambda_{\pm}\left(q\right)}\left(e^{2i\omega_{1}t}+1\right)-ib_{\Lambda_{\pm}\left(q\right)}\left(e^{2i\omega_{1}t}-1\right)\right]/2$,
and comparing to Eq. (\ref{eq:line_delta_rho_sol}), we obtain a correspondence
between $\Lambda$$\left(q\right)$ and s$\left(q\right)$ giving
the plasmon dispersion close to the resonance:
\begin{equation}
\Lambda_{\pm}\left(q\right)=is_{\pm}\left(q\right)+\omega_{1}.\label{eq:mu_quasienergy}
\end{equation}
Note that the Floquet exponent $\Lambda\left(q\right)$ is defined
modulo $2\omega_{1}$ (because a change by $2\omega_{1}$ can always
be absorbed into $u_{\Lambda}\left(t\right)$). In analogy to the
Bloch theory of electronic band structure, $2\omega_{1}$ plays the
role of a reciprocal lattice vector. 

\begin{figure}
\centering{}\includegraphics[width=0.95\columnwidth]{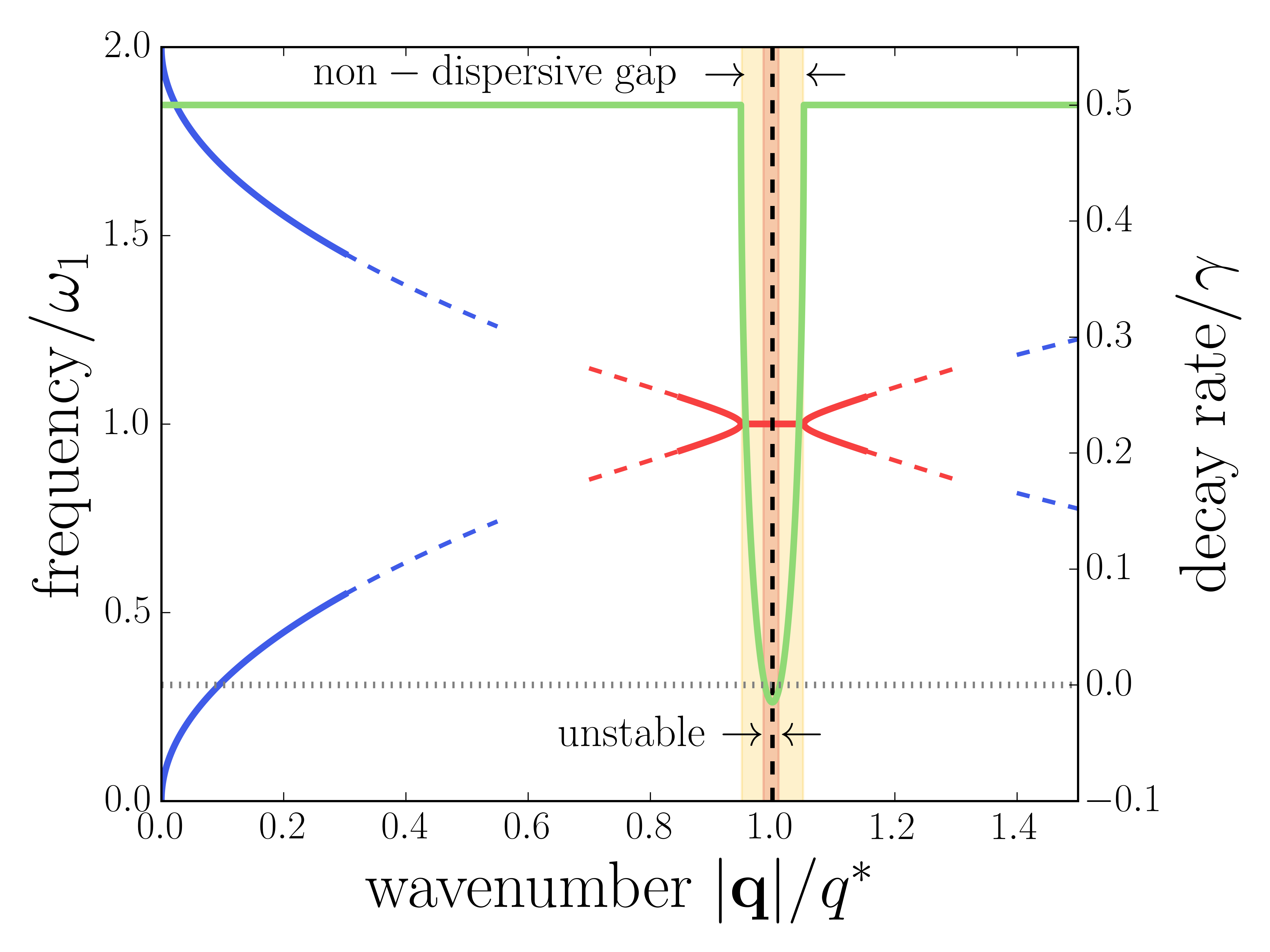}\caption{Onset of the plasmon instability induced by modulated Floquet parametric
driving with the modulation frequency $2\omega_{1}$. Red and blue
lines show the the quasi-energy dispersion $\mathrm{Re}\left[\Lambda\left(q\right)\right]$
of parametrically driven plasmons (see Eq. (\ref{eq:mu_quasienergy})
and discussion below). Far away from the critical $q^{*}$, 2D plasmons
retain the characteristic square-root shape of their dispersion relation
(blue lines). The quasi-energy dispersion is invariant with respect
to shifts by $2\omega_{1}$ -- the reciprocal lattice constant in
frequency space. Near $q^{*}$, where $\omega_{\mathrm{pl}}\left(q^{*}\right)\approx\omega_{1}$
holds, the dispersion is strongly altered by the driving (red lines).
A gap which hosts non-dispersive modes opens around $q^{*}$ (yellow
shading). Merging branches of the dispersion at the edges of the non-dispersive
gap indicate exceptional points with diverging group velocities. We
chose the modulation amplitude $h$ to lie slightly above the instability
threshold: damping is negative in a small interval around $q^{*}$,
leading to an exponential growth of unstable modes (red shading).
However, exceptional points and non-dispersive states appear even
for subcritical driving strenghts.\label{fig:Onset-of-the_instability}}
\end{figure}

While Eqs. (\ref{eq:s}) and (\ref{eq:mu_quasienergy}) are valid
near $\omega_{\mathrm{pl}}\left(q\right)=\omega_{1}$, for small $q$,
keeping higher order derivatives neglected above in the slowly varying
envelope approximation, and ignoring damping, we find that the two
branches of the plasmon dispersion are given by $\Lambda_{+}\left(q\right)\approx2\omega_{1}-\omega_{\mathrm{pl}}\left(q\right)/\sqrt{1+h^{2}/4}$
and $\Lambda_{-}\left(q\right)\approx\omega_{\mathrm{pl}}\left(q\right)/\sqrt{1+h^{2}/4}$.
For the $\Lambda_{-}$ branch, we find $b_{\Lambda_{-}\left(q=0\right)}=ia_{\Lambda_{-}\left(q=0\right)}$,
giving $\delta\rho_{q}\approx a_{\Lambda_{-}\left(q=0\right)}e^{i\omega_{\mathrm{pl}}\left(q\right)t/\sqrt{1+h^{2}/4}}$.
Therefore, the $\Lambda_{-}$ branch describes a counter-propagating
mode with the frequency $-i\omega_{\mathrm{pl}}\left(q\right)/\sqrt{1+h^{2}/4}$,
lifted by one reciprocal lattice vector $2\omega_{1}$. We conclude
that, away from the resonance, the parametric driving influences the
plasmons only weakly. The plasmon dispersion in the presence of parametric
diving $\Lambda\left(q\right)$ is shown in Fig. \ref{fig:Onset-of-the_instability}.

Exceptional points appear at wavenumbers $q_{\mathrm{exc}}$, where
the $\pm$-branches of $s_{\pm}\left(q\right)$ merge, i.e., when
the condition $\omega_{\mathrm{pl}}^{2}\left(q_{\mathrm{exc}}\right)=\omega_{1}^{2}\left(1\pm h/2\right)$
is met. This results in a diverging group velocity $\partial s_{\pm}\left(q\right)/\partial q\sim\left|q-q_{\mathrm{exc}}\right|^{-1/2}$
for $q\rightarrow q_{\mathrm{exc}}$. In the interval between the
two exceptional points $q_{\mathrm{exc}}$, $s_{\pm}\left(q\right)$
is purely real. This corresponds to non-dispersive plasmons, oscillating
at frequency $\omega_{1}$ (see Fig. \ref{fig:Onset-of-the_instability}).

Let us now consider the stability of plasmon modes under MFPD. The
instability condition $\mathrm{Re}\left(s\right)>0$ is realized for
$h>2\gamma/\omega_{1}$ in a narrow frequency range around $\omega_{\mathrm{pl}}\left(q\right)=\omega_{1}$
given by 
\begin{equation}
\left[\omega_{\mathrm{pl}}^{2}\left(q\right)-\omega_{1}^{2}\right]^{2}<\omega_{1}^{4}h^{2}/4-\omega_{1}^{2}\gamma^{2}.\label{eq:unstable_region}
\end{equation}
The fastest growing modes have wavenumbers $q^{*}$ determined by
the condition
\begin{equation}
\pm\omega_{\mathrm{pl}}\left(q^{*}\right)=\omega_{1}.\label{eq:q_star_def}
\end{equation}
Thus, the parametric driving will excite plasmons with the frequency
$\omega_{1}$, whose amplitude will grow according to $\text{\ensuremath{\delta\rho}}_{q^{*}}\sim e^{\left(h\omega_{1}/4-\gamma/2\right)t}$.
Notice that the system's response breaks the discrete time translation
symmetry of the drive with respect to translations by $T=\pi/\omega_{1}$.

The periodic driving opens a vertical non-dispersive gap around $q^{*},$
where the two branches of the plasmon dispersion meet. Inside the
gap, the real part of $\Lambda_{\pm}\left(q\right)$ is flat. This
phenomenon is known from photonic time varying media and photonic
time crystals, where it is often referred to as momentum-gap or k-gap
\citep{galiffi2022photonics_time_varying_ptc,lyubarov2022photonic_time_crystal_amplified}.
In fact, Eq. (\ref{eq:param_plasmon_driving}) maps onto the equation
describing the propagation of light through a photonic time crystal
\citep{lyubarov2022photonic_time_crystal_amplified}. The full plasmon
dispersion near the onset of the instability ($h\omega_{1}/4\gtrsim\gamma/2$)
is illustrated in Fig. \ref{fig:Onset-of-the_instability}.

\subsection{Enhancing plasmonic quality factors}

\begin{figure}
\centering{}\includegraphics[width=0.95\columnwidth]{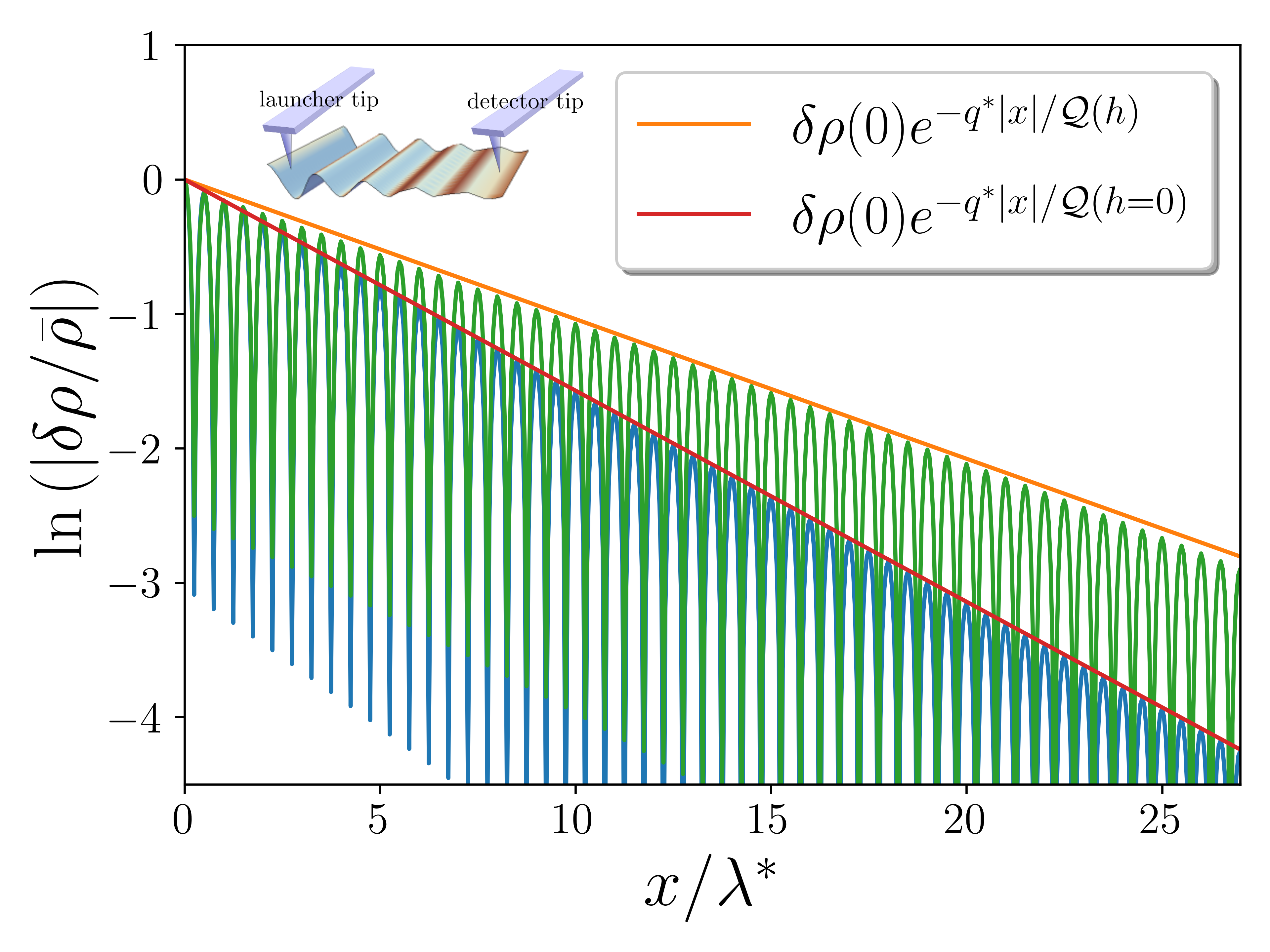}\caption{Enhancement of the plasmon quality factor by MFPD (logarithmic plot).
A source at $x=0$ induces plasmons propagating along the $x$-axis.
The plasmon amplitude is exponentially damped according to Eq. (\ref{eq:rho_decay}).
Blue and green curves show the plasmon wave with MFPD switched on
and off, respectively. The enhancement of the plasmon quality factor
shows in a larger propagation length and agrees well with the prediction
of Eq. (\ref{eq:quality_factor_h}). A driving strength of $h=0.75h_{c}$
was used in the simulation.\label{fig:Enhancement-of-the_quality}}
\end{figure}

In this section, we show how MFPD below the instability threshold
can be used to enhance plasmonic quality factors. As an example we
consider a plasmon source that is located at $x=0$ and is, for simplicity,
uniform along the $y$-axis, which is a reasonable approximation for
a strip whose width is shorter than the plasmon wavelength \citep{sun2022graphene_entangled_plasmon_pairs}.
This source is placed into an MFPD driven plasmonic medium. We use
Eq. (\ref{eq:param_plasmon_driving}) with a pointlike source term
oscillating at frequency $\omega_{1}$ added to the left hand side
to analyze plasmon propagation in this setting. In the Supplementary
Information, we show that for any fixed $t\rightarrow\infty$, the
plasmon amplitude away from the source behaves as
\begin{equation}
\delta\rho\left(x\right)\sim\delta\rho\left(0\right)e^{-\frac{q^{*}}{\mathcal{Q}\left(h\right)}\left|x\right|},\label{eq:rho_decay}
\end{equation}
where 
\begin{equation}
\mathcal{Q}\left(h\right)=\frac{1}{\sqrt{\gamma^{2}/\omega_{1}^{2}-h^{2}/4}}\label{eq:quality_factor_h}
\end{equation}
is the plasmon quality factor. In the absence of MFPD, $\mathcal{Q}\left(h\right)$
reduces to the well known expression $\mathcal{Q}\left(h=0\right)=\omega_{1}/\gamma$.
Eq. (\ref{eq:quality_factor_h}), shows that for driving strengths
$h$ close to, but below the instability threshold $h_{c}=2\gamma/\omega_{1},$
the quality of the plasmon resonance can be strongly enhanced. To
demonstrate the enhancement, we perform a numerical simulation of
Eq. (\ref{eq:param_plasmon_driving}). For all simulations in this
paper we use the Dedalus spectral solver \citep{burns2020dedalus}.
The results (see Fig. \ref{fig:Enhancement-of-the_quality}) show
a good agreement with Eq. (\ref{eq:rho_decay}), and demonstrate that
a significant enhancement of $\mathcal{Q}$ can be achieved at driving
strengths below the instability threshold. This opens the possibility
of enhancing the quality of plasmon resonances with high frequency
optical drives.

Finally, we note that the quality factor $\mathcal{Q}$ depends on
temperature. In experiments with graphene plasmons, $\mathcal{Q}$
roughly decreases by a factor of two when the temperature is raised
from $100\,\mathrm{K}$ to $200\,\mathrm{K}$ \citep{ni2018_plasmon_quality_factor_graphene}.
Below, we estimate that the temperature raise induced by MFPD is on
the order of a few tens of Kelvin for $h=h_{c}$. Thus MFPD can compensate
for the heating induced lowering of $\mathcal{Q}$, if $h$ is chosen
close to $h_{c}$.

\subsection{Crystallization\label{subsec:Crystallization}}

We now turn to describing pattern formation that occurs for MFPD above
the critical driving threshold, $h>h_{c}$. Having identified the
wavenumber $q^{*}$ of the unstable modes in Eqs. (\ref{eq:plasmon_dispersion})
and (\ref{eq:q_star_def}), we now study the steady state spatial
structure of the plasmon charge density once the initial exponential
growth has been saturated by the nonlinearities. We use the nonlinear
equation of motion for plasmons with an oscillating mass term, Eq.
(\ref{eq:rho_nonlin_eq}), derived in Sec. \ref{subsec:Oscillating-mass}.
For simplicity, we focus on a quasi 1D strip whose width along the
$y$-direction is smaller than the plasmon wavelength $\lambda^{*}=2\pi/q^{*}$.
These boundary conditions are implemented by setting $\rho=0$ outside
the strip.

Parametrically driven nonlinear waves were thoroughly studied in nonlinear
spin and optical systems in the context of wave turbulence \citep{zakharov1975spin_wave_turbulence}.
In general, the saturation of unstable modes can be described within
the subspace of modes that satisfy the condition $2\omega_{1}=\omega_{\mathrm{pl}}\left(q\right)+\omega_{\mathrm{pl}}\left(-q\right)$
\citep{zakharov1975spin_wave_turbulence}. For the strip geometry,
where $q_{y}$ is fixed to zero, these are precisely the linearly
unstable modes of Eqs. (\ref{eq:line_delta_rho_sol}) and (\ref{eq:slowly_varying_ansatz}).
Physically, the above condition reflects that the uniform parametric
driving cannot supply momentum to the system, and plasmons can only
be created in pairs with wavenumbers $q_{x}=\pm q^{*}$. Thus, the
saturation to the final steady state will be captured by the ansatz
\begin{align}
\delta\rho_{s}\left(t,y\right) & =a\left(t\right)\cos\left(\omega_{1}t\right)\cos\left(q^{*}\cdot x\right)\nonumber \\
 & \quad+b\left(t\right)\sin\left(\omega_{1}t\right)\cos\left(q^{*}\cdot x\right).\label{eq:delta_rho_ansatz}
\end{align}
A similar instability appears in parametrically driven shallow water
waves, so called Faraday waves \citep{faraday1831_patterns,benjamin1954_faraday_stability,kumar1994_faraday,edwards1994patterns_faraday,muller1994_quasipatterns_faraday,chen_vinals_1997_faraday_pattern_prl,chen_vinals_1999_faraday_pattern_sel}.
The Faraday instability has also been reported for cold atom systems
\citep{di2023instabilitiesCold_atoms_faraday_instability_experiment,dupont2022Cold_atoms_faraday_instability}
and Luttinger liquids \citep{fazzini2021Luttinger_floquet_charge_density_wave}.

As shown in Sec. \ref{subsec:Analysis-of-nonlinear_pattern} inserting
the ansatz of Eq. (\ref{eq:delta_rho_ansatz}) into the plasmon equation
of motion (\ref{eq:rho_nonlin_eq}) leads to the set of amplitude
equations
\begin{align}
\dot{a} & =-\alpha b-\frac{1}{2}\gamma a+\beta b\left(a^{2}+b^{2}\right)\nonumber \\
\dot{b} & =-\alpha a-\frac{1}{2}\gamma b-\beta a\left(a^{2}+b^{2}\right),\label{eq:a_b_nonline_eom}
\end{align}
with $\alpha=\omega_{1}h/4$ and $\beta=\omega_{1}/32\left(\bar{\rho}\right)^{2}$.
To gain an intuition for the steady state, it is useful to consider
the fixed point, where $\dot{a}\left(a_{\mathrm{fp}},b_{\mathrm{fp}}\right)=\dot{b}\left(a_{\mathrm{fp}},b_{\mathrm{fp}}\right)=0$
and solve for the amplitude $r_{\mathrm{fp}}^{2}=a_{\mathrm{fp}}^{2}+b_{\mathrm{fp}}^{2}$.
Here, the subscript ``fp'' stands for values at the fixed point.
One finds
\begin{equation}
r_{\mathrm{fp}}^{2}=a_{\mathrm{fp}}^{2}+b_{\mathrm{fp}}^{2}=\frac{\alpha}{\beta}\sqrt{\left[1-\left(\frac{\gamma}{2\alpha}\right)^{2}\right]}.\label{eq:fixed_point_r}
\end{equation}
This expression indicates a critical point at $h_{c}=2\gamma/\omega_{1}$,
beyond which, for $h>h_{c}$, physically meaningful fixed points with
a real, finite amplitude $r_{\mathrm{fp}}$ exist. Equation (\ref{eq:fixed_point_r}),
also shows that the amplitude $r_{\mathrm{fp}}$ has two distinct
scaling regimes connected by a crossover. For large $h$, the damping
$\gamma$ is negligible and $r_{\mathrm{fp}}\approx\sqrt{\alpha/\beta}$.
On the other hand, close to the critical point $h_{c}$, we find $r_{\mathrm{fp}}\approx\sqrt{\alpha_{c}/\beta}\left(2\epsilon\right)^{1/4}$,
where $\epsilon=\left(\alpha-\alpha_{c}\right)/\alpha_{c}$ with $\alpha=\omega_{1}h_{c}/4$.
We will refer to these two cases as the weakly damped and the strongly
damped, respectively. In what follows, we use Eqs. (\ref{eq:a_b_nonline_eom})
to study the transition to the crystalline non-equilibrium steady
state in both cases.

\subsubsection{Transition in the weakly damped case}

\begin{figure}
\centering{}\includegraphics[scale=0.8]{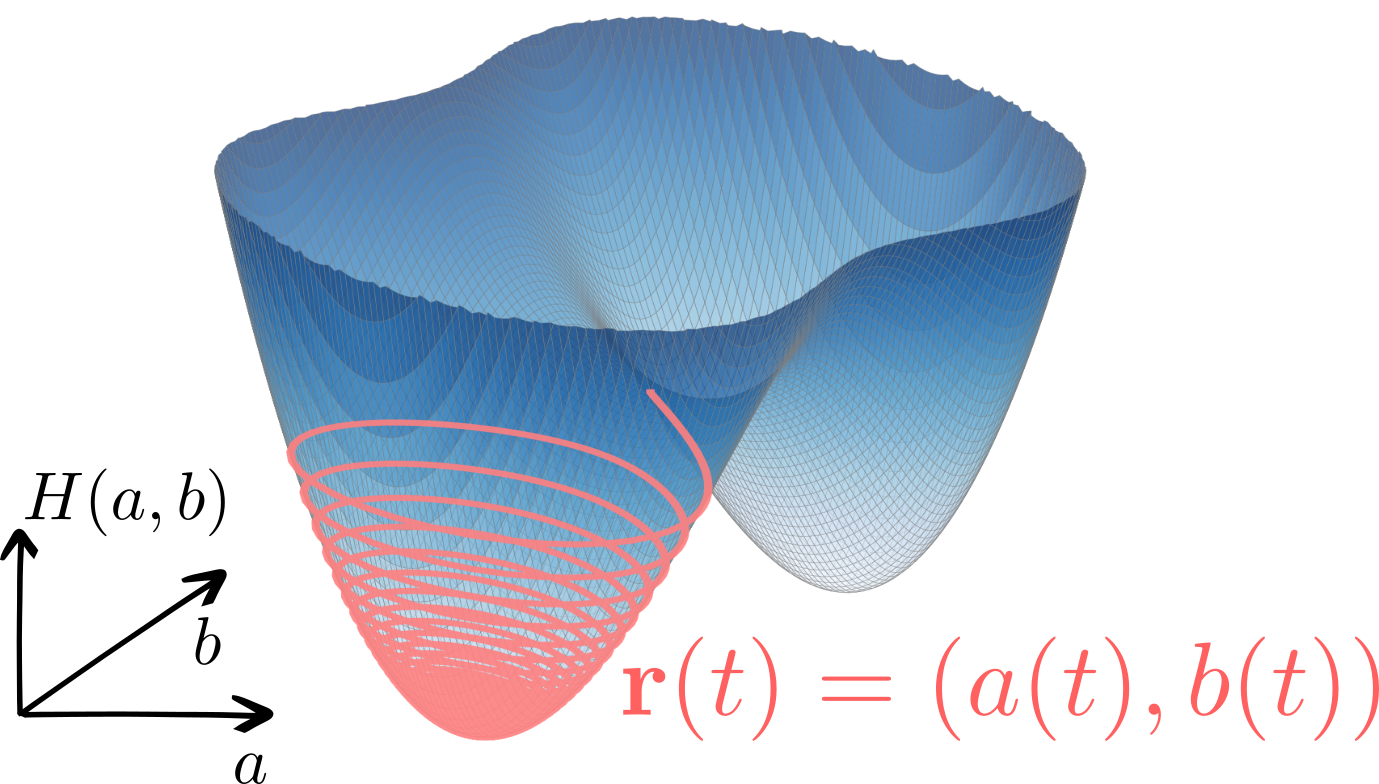}\caption{The effective Hamiltonian of Eq. (\ref{eq:eff_hamiltonian}) in amplitude
space, with the two minima corresponding to symmetry breaking states.
The red trajectory is a solution to Eqs. (\ref{eq:a_b_nonline_eom})
and shows the transition to the crystalline state in the presence
of damping.\label{fig:effective-Hamiltonian}}
\end{figure}

For $\gamma=0$, the Eqs. (\ref{eq:a_b_nonline_eom}) correspond to
the equations of motion arising from the Hamiltonian
\begin{equation}
H\left(a,b\right)=\frac{1}{2}\alpha\left(a^{2}-b^{2}\right)+\frac{1}{4}\beta\left(a^{4}+b^{4}\right)+\frac{1}{2}\beta a^{2}b^{2}.\label{eq:eff_hamiltonian}
\end{equation}
Similar Hamiltonians are obtained for nonlinear parametric oscillators
in both quantum and classical limits \citep{dykman1998fluctuational_transitions_parametric_osc,marthaler2006parametric_osc_quantum_switching}.
The stable minima of the Hamiltonian (\ref{eq:eff_hamiltonian}) are
located at:
\begin{align}
a_{*} & =0,\ b_{*}=\pm\sqrt{\frac{\alpha}{\beta}},\label{eq:a_b_sol}
\end{align}
In the presence of weak damping, Eqs. (\ref{eq:a_b_nonline_eom})
predict that, for any initial condition, the trajectory $r\left(t\right)=\left(a\left(t\right),b\left(t\right)\right)$
will descend to one of the minima predicted by Eq. (\ref{eq:a_b_sol})
(see Fig. \ref{fig:effective-Hamiltonian}). The steady state will
be a standing wave of the form of Eq. (\ref{eq:delta_rho_ansatz})
with the amplitude given by Eq. (\ref{eq:a_b_sol}). 

We note that noise can significantly alter the dynamics of the plasmon
modes. It can, e.g., lead to transitions between the two minima of
the Hamiltonian (\ref{eq:eff_hamiltonian}) \citep{dykman1998fluctuational_transitions_parametric_osc}.
In the case of multiple non-degenerate minima noise can determine
the ultimate stable configuration \citep{dykman1998fluctuational_transitions_parametric_osc}.

\subsubsection{Transition in the strongly damped case}

\begin{figure}
\centering{}\includegraphics[width=0.95\columnwidth]{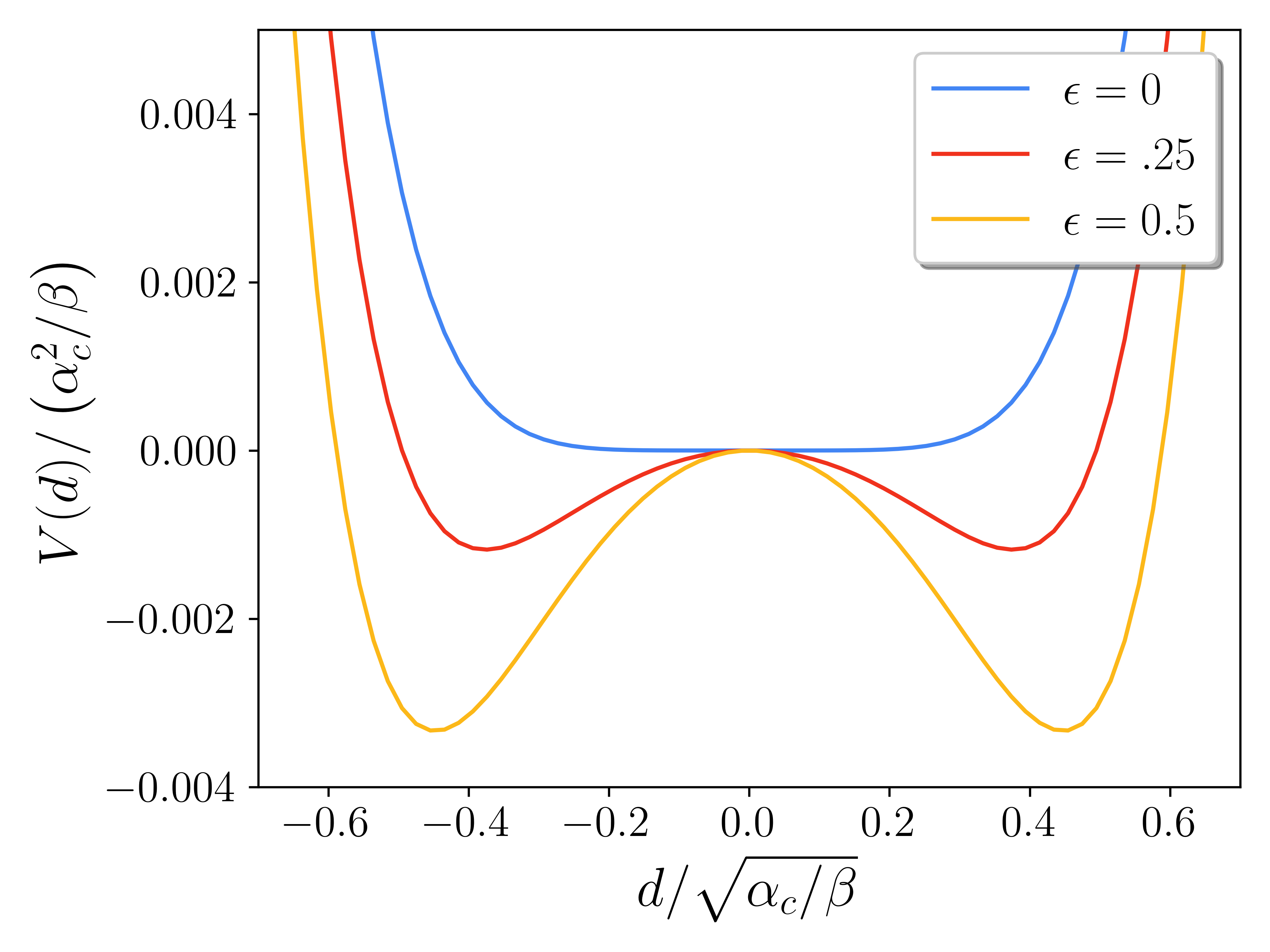}\caption{The potential $V\left(d\right)$ of Eq. (\ref{eq:symmetry_breaking_potential})
with the two minima corresponding to symmetry breaking steady states
of the driven systems. The system undergoes a continuous symmetry
breaking transition when driven above the threshold value $h_{c}=2\gamma/\omega_{1}$,
i.e. when $\epsilon=h/h_{c}-1>0$. \label{fig:double_well}}
\end{figure}

Next, we address the case where $h$ is only slightly above the instability
threshold $h_{c}=2\gamma/\omega_{1}$. Expanding Eqs. (\ref{eq:a_b_nonline_eom})
in $\epsilon=\left(h-h_{c}\right)/h_{c}$ (see Supplementary Information),
we find that $a\left(t\right)\approx-b\left(t\right)$ and the time
dependence of the mode amplitudes is -- to first order in $\epsilon$
-- captured by the equation $\dot{d}=-\partial V\left(d\right)/\partial d$,
where $d=a-b$ and the potential $V\left(d\right)$ is given by
\begin{equation}
V\left(d\right)=-\epsilon\frac{\alpha_{c}}{2}d^{2}+\frac{\beta^{2}}{48\alpha_{c}}d^{6}.\label{eq:symmetry_breaking_potential}
\end{equation}
The equation for $\dot{d}$ describes a gradient descent dynamics
towards one of the two minima of the potential (\ref{eq:symmetry_breaking_potential}).
We plot the effective potential for different values of $\epsilon$
in Fig. \ref{fig:double_well} and conclude that the transition into
the symmetry breaking state is continuous, i.e., the system goes through
a soft bifurcation when $h$ reaches the critical driving strength
$h_{c}$. In terms of the original $a$ and $b$, the potential minima
{[}and approximate fixed points of Eqs. (\ref{eq:a_b_nonline_eom}){]}
are located at
\begin{equation}
a=-b=\pm\sqrt{\frac{\alpha_{c}}{\beta}}\left(\epsilon/2\right)^{1/4}.\label{eq:close_threshold_fixed_points}
\end{equation}

\subsubsection{Crystallization in 2D}

\begin{figure}
\centering{}\includegraphics[width=0.2\paperwidth]{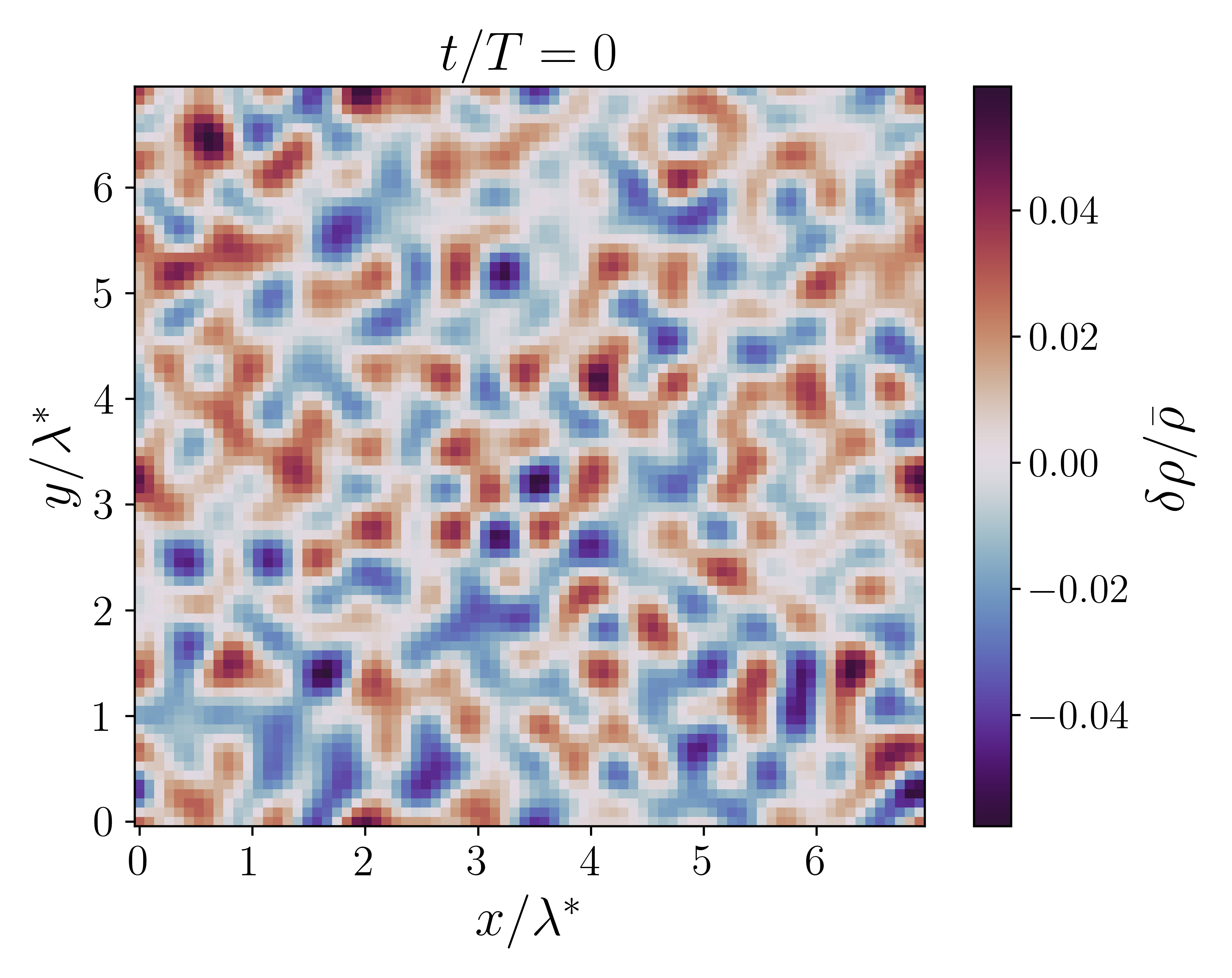}\includegraphics[width=0.2\paperwidth]{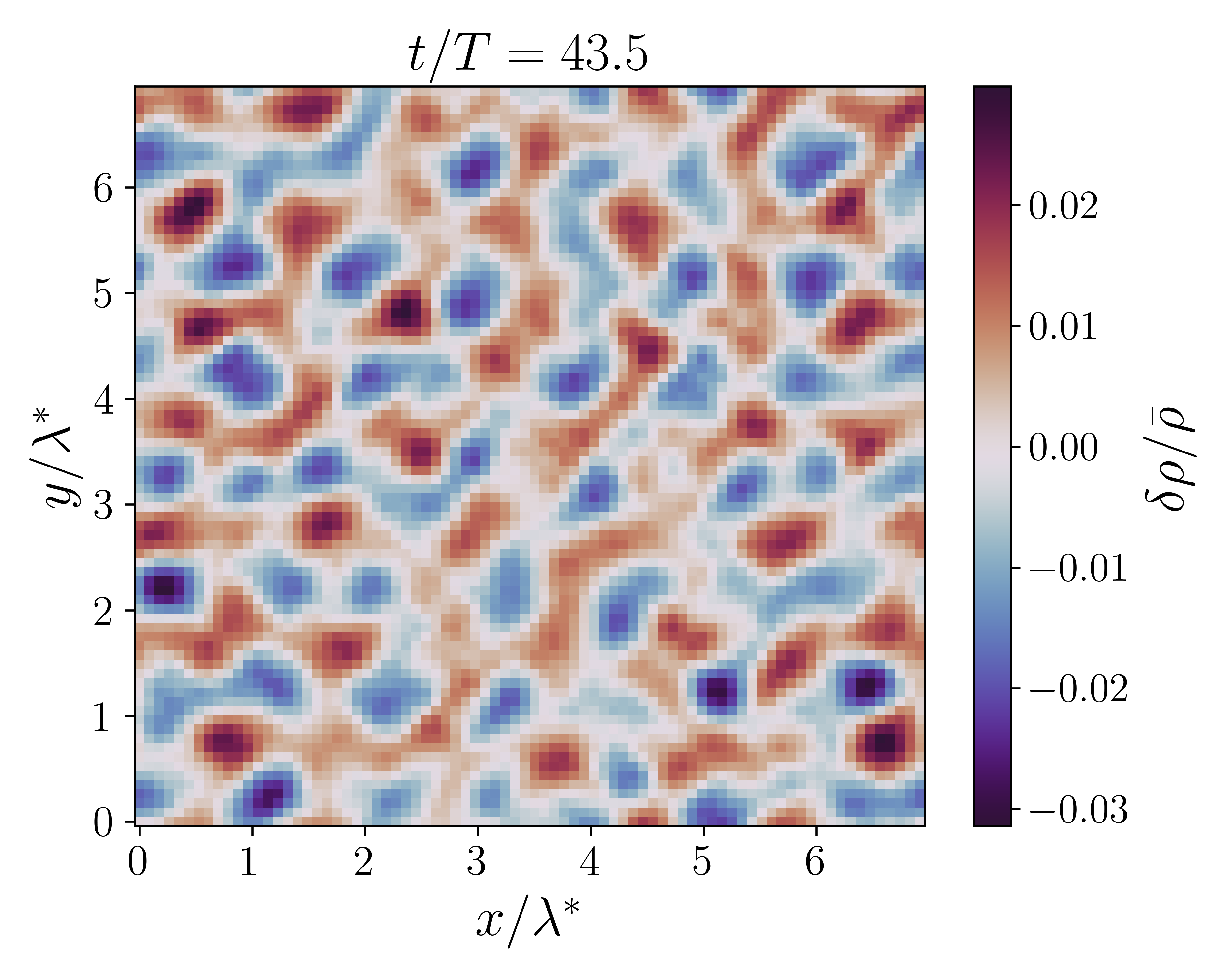}\\
\includegraphics[width=0.2\paperwidth]{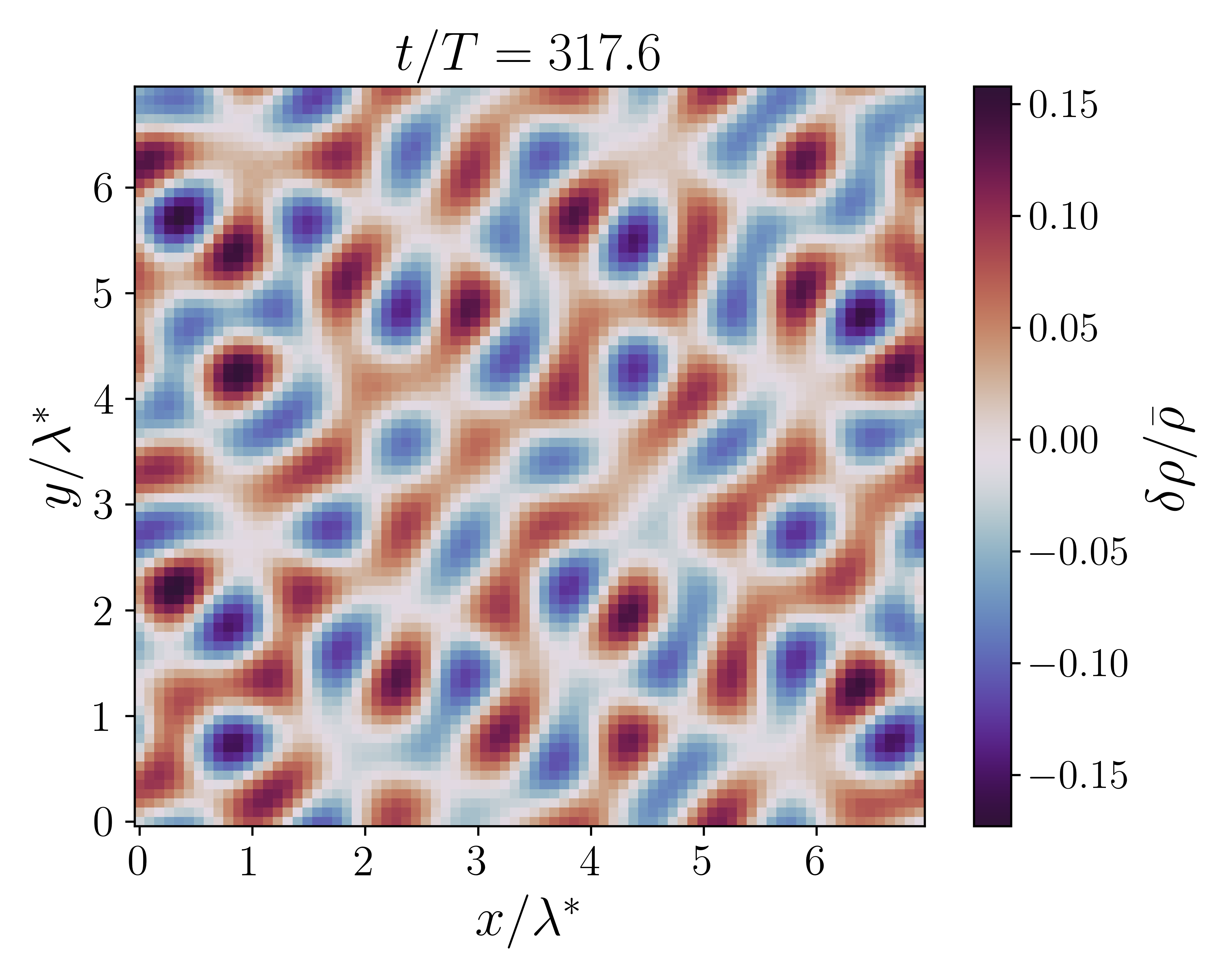}\includegraphics[width=0.2\paperwidth]{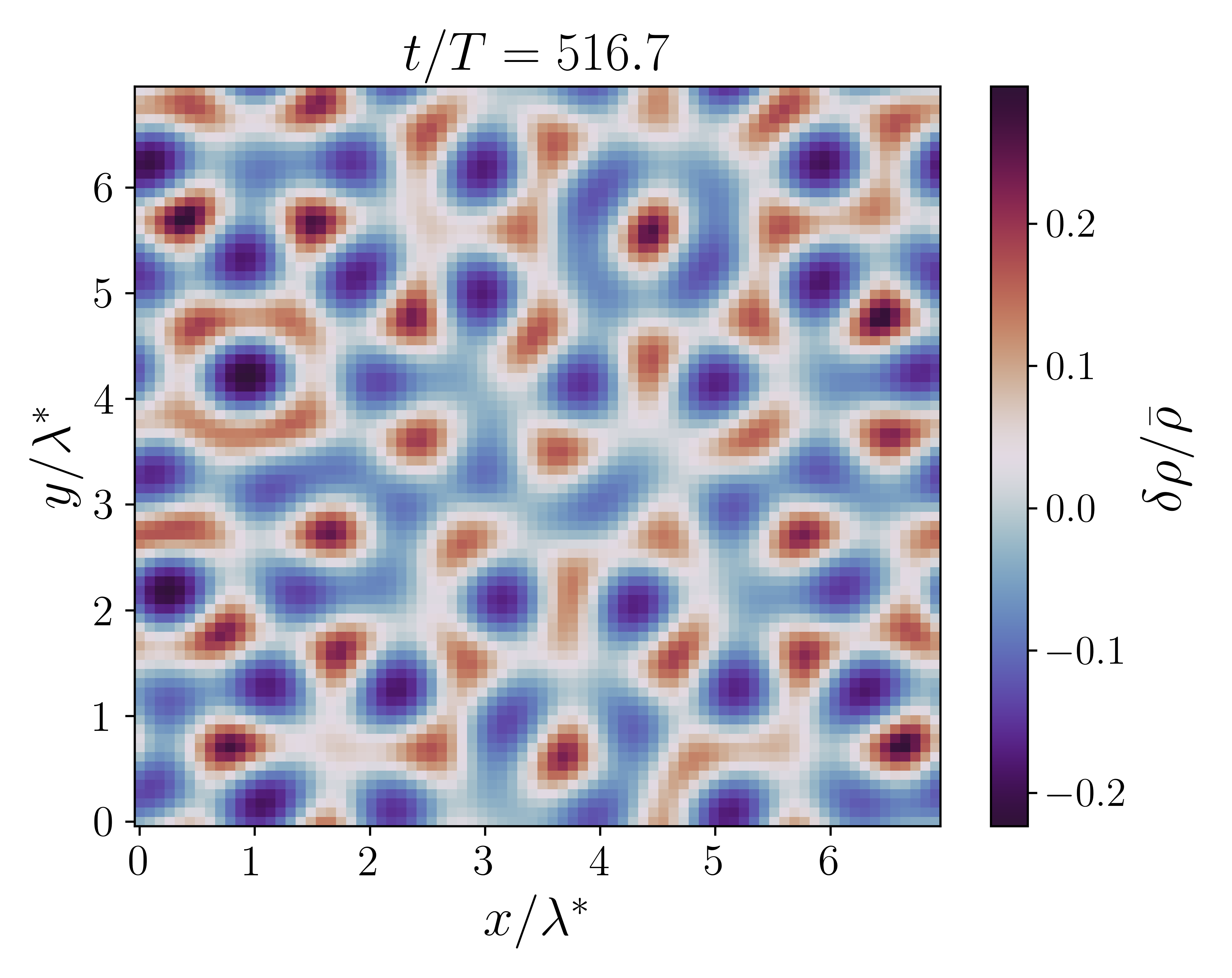}\\
\includegraphics[width=0.2\paperwidth]{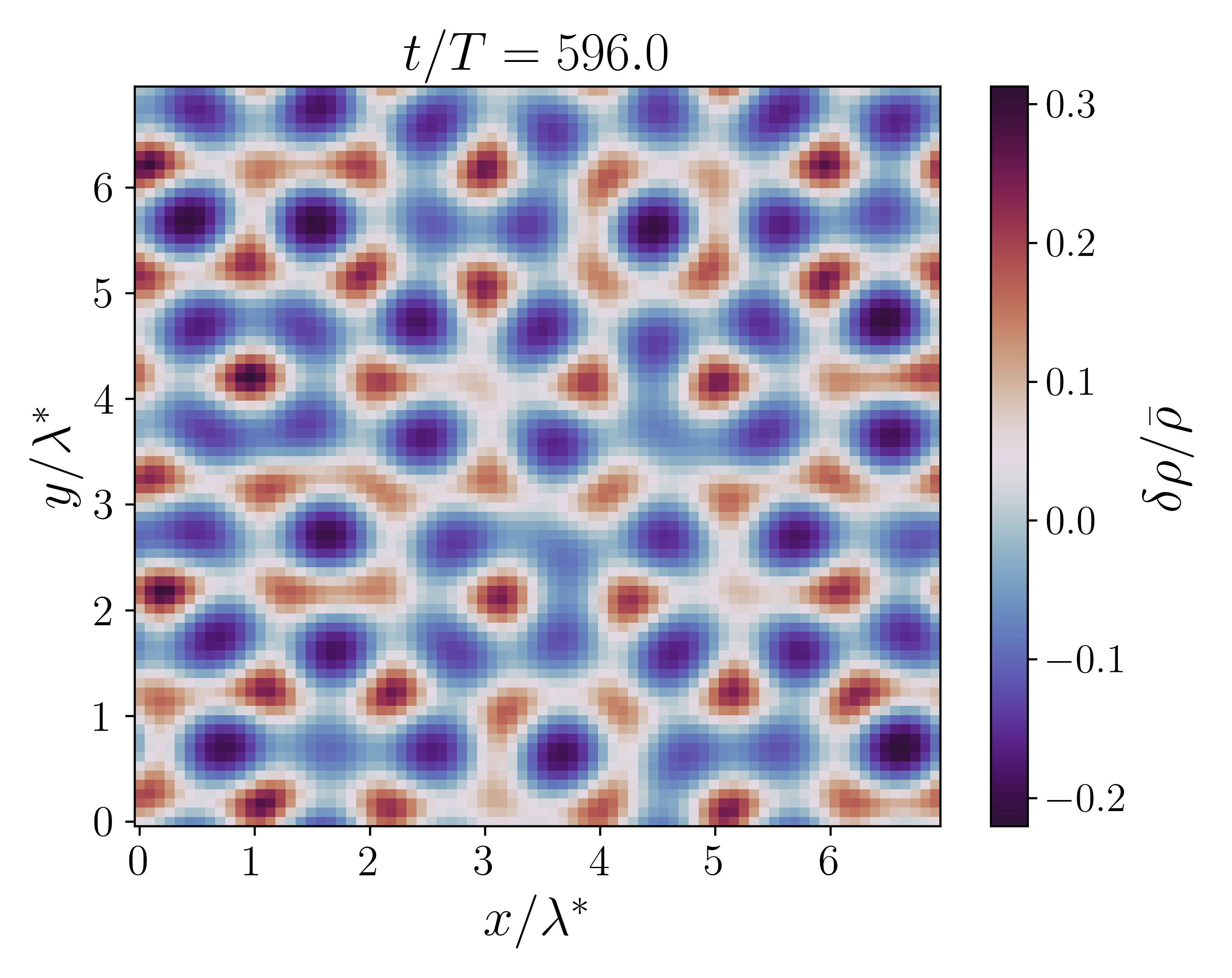}\includegraphics[width=0.2\paperwidth]{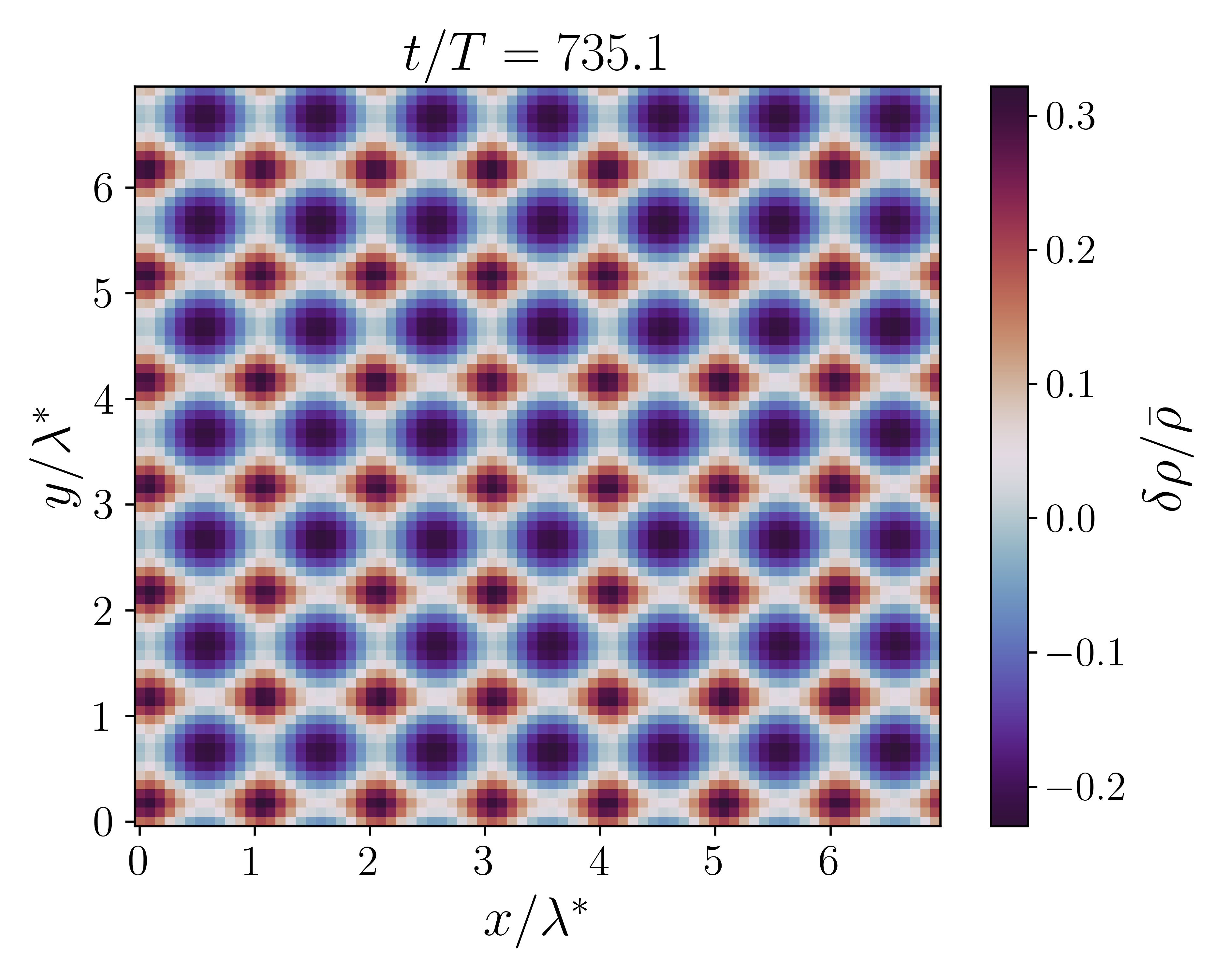}\caption{Plasmonic pattern formation under MFPD in two dimensions in the strongly
damped regime. The plots show a numerical solution of Eqs. (\ref{eq:conti})-(\ref{eq:poisson_eq}).
Commensurate periodic boundary conditions were used in both directions.
The driving and damping parameters used in this plot are $h=0.05$,
$\gamma=0.045\omega_{1}$. An initial patterns with a characteristic
periodicity of $\lambda^{*}=2\pi/q^{*}$ forms quickly from the random
initial conditions. The final stable square pattern is reached after
a series of transformations. Time is given in units of $T=\pi/\omega_{1}$.
\label{fig:sim_patterns}}
\end{figure}

After studying plasmon pattern formation in 1D, we extend our results
to two dimensions, focusing on the a driving strength slightly above
the instability threshold. In two dimensions, the condition $2\omega_{1}=\omega_{\mathrm{pl}}\left(q\right)+\omega_{\mathrm{pl}}\left(-q\right)$
for selecting modes that contribute to the steady-state solution does
not specify a finite set of modes. Instead, all modes with wavevectors
$\mathbf{q}^{*}$ that satisfy $\left|\mathbf{q}^{*}\right|=q^{*}$
can, in principle, contribute. Studies of pattern formation in Faraday
waves \citep{chen_vinals_1999_faraday_pattern_sel,chen_vinals_1997_faraday_pattern_prl,muller1994_quasipatterns_faraday}
suggest that a finite number of modes $N$ with wavevectors $\mathbf{q}_{i}^{*}$
will be selected, such that the angle between the wavevectors is given
by $\pi/N$. While for $N=1$, one obtains a standing wave, the $N=2$
case corresponds to a square tiling and the $N=3$ case gives a hexagonal
or triangular tiling. To find the plasmon pattern to which the system
converges in the steady state, we performed numerical simulations
of Eqs. (\ref{eq:conti})-(\ref{eq:nav_stokes}) under MFPD (for details
we refer to the Supplementary Information). As closest feasible approximation
to an infinite plane, we choose a periodic domain with side length
commensurate with the plasmon wavelength $\lambda^{*}=2\pi/q^{*}$.
Curiously, the patterns quickly reach states with complex, quasiperiodic
patterns of waves with wavelengths $\lambda^{*}$, then slowly settle
on a nearly square pattern after a series of transformations. A typical
cristallization process is shown in Fig. \ref{fig:sim_patterns}.

We find that, generally, plasmons arrange themselves in square patterns
in the steady state. This is even true for square domains incommensurate
with $\lambda^{*}$. We note that the observed patterns are are not
perfect square tilings. This is due to the presence of small amplitude
modes with larger wavenumbers caused by the nonlinearities (see Supplementary
Information). By fine tuning the boundary conditions, we were also
able to realize a triangular pattern (see Supplementary Information).
We also note that, under experimental conditions, noise can influence
the steady state to which the system finally settles \citep{dykman1998fluctuational_transitions_parametric_osc}.
Our conclusion is that while a square tiling is preferred generally,
boundary conditions can affect the final pattern.

\subsection{Goldstone-like phonons\label{subsec:Goldstone-like-phonons}}

The crystalline steady state breaks the translational symmetry of
the system. We show here that this breaking of a continuous symmetry
manifests itself in the presence of Goldstone-like phonon modes. We
focus on the strip geometry.

Above, we found the steady state solution: $\delta\rho=\pm\sqrt{\alpha/\beta}\sin\left(\omega_{1}t\right)\cos\left(q^{*}x\right)$
{[}see Eqs. (\ref{eq:delta_rho_ansatz}) and (\ref{eq:a_b_sol}){]}
for the weakly damped case, which we will consider in this section.
This solution remains valid if we shift the phase of the spatial part
by $\phi$:
\begin{equation}
\delta\rho=\pm\sqrt{\frac{\alpha}{\beta}}\sin\left(\omega_{1}t\right)\cos\left(q^{*}x+\phi\right).\label{eq:sol_phaseshift}
\end{equation}
If homogeneous and static, the phase shift can always be eliminated
by the coordinate transformation $x\rightarrow x-\phi/q^{*}$. However,
if the phase shift is spatially dependent, Eq. (\ref{eq:rho_nonlin_eq})
determines its dynamics. Thus, we investigate the dynamics of $\phi\left(t,x\right)$,
via the ansatz $\delta\rho=\pm\sqrt{\alpha/\beta}\sin\left(\omega_{1}t\right)\cos\left[q^{*}x+\phi\left(t,x\right)\right]$,
which remains an approximate solution to Eq. (\ref{eq:rho_nonlin_eq}).
Since a uniform $\text{\ensuremath{\phi}}$ has no influence on the
dynamics of the system, we expect slow dynamics for a long wavelength
spatial dependence of $\phi\left(t,x\right)$, i.e., we expect Goldstone-like
modes.

In Sec. \ref{subsec:Deriving-the-Goldstone} we find that the dynamics
of $\phi\left(t,x\right)$ is governed by a wave equation that, to
linear order, is solved by plane waves $\phi\left(t,x\right)=\mathrm{Re}\left[e^{-i\Omega_{\pm}\left(Q\right)t+iQx}\right]$
with the dispersion:
\begin{equation}
\Omega_{\pm}\left(Q\right)\approx\pm\frac{\omega_{1}}{q^{*}}\sqrt{\left(1+3h\right)Q^{2}}.\label{eq:Goldstone_dispersion}
\end{equation}
We note that similar modes have been observed experimentally in parametrically
driven classical liquids \citep{domino2016faraday_phonons}. Interestingly,
the dispersion (\ref{eq:Goldstone_dispersion}) is linear, unlike
the plasmon plasmon dispersion $\omega_{\mathrm{pl}}\left(q\right)$.
On long scales, the lattice distortions are net-neutral, therefore
lacking a long-range restoring force.

\subsubsection{Optical modes}

Besides the Goldstone modes, the symmetry breaking state supports
optical modes. These correspond to oscillations of the amplitudes
$a$, $b$ in Eq. (\ref{eq:a_b_nonline_eom}).

To derive the dispersion of the optical modes, we expand the effective
Hamiltonian in Eq. (\ref{eq:eff_hamiltonian}) around the minimum
at $a=0$, $b=\sqrt{\alpha/\beta}$:
\begin{equation}
H\approx-\frac{\alpha^{2}}{4\beta}+\alpha\left(\delta b^{2}+\delta a^{2}\right).
\end{equation}
The corresponding Hamilton's equations read

\begin{align}
\delta\dot{b} & =2\alpha\delta a\nonumber \\
\delta\dot{a} & =-2\alpha\delta b,
\end{align}
and are solved by a uniform oscillation of the amplitudes with the
frequency $2\alpha$:
\begin{equation}
\delta\ddot{a}+4\alpha^{2}\delta a=0,\label{eq:optical_mode_eq}
\end{equation}
with $\alpha=\omega_{1}h/4$.

The collective modes of the symmetry broken state described by Eqs.
(\ref{eq:Goldstone_dispersion}) and (\ref{eq:optical_mode_eq}) are
one possible experimental signature of the crystalline state. When
the dissipation rate $\gamma$ is included, the modes will obtain
a negative imaginary part $\sim-i\gamma$, broadening the resonances.

\subsection{Heating and Landau damping\label{subsec:Heating-and-Landau}}

Coherent driving of interacting systems leads to heating \citep{lazarides2014_floquet_heating,dalessio2014_floquet_heating,bukov2016heating_floquet,bukov2015_floquet_heating_prethermal,else2017_floquet_heating_prethermal,mori2018_floquet_heating_prethermal,reitter2017_floquet_heating_experiment,singh2019_floquet_heating_experiment,galitsky1970_driven_semiconductor_steady_state,shirai2015condition_for_gibbs_in_floquet_steady_state,seetharam2015baths_controlled_floquet_population,iadecola2015occupation_topo_floquet,seetharam2019floquet_steady_state_interacting,liu2015classification_of_floquet_statistical_distribution,abanin2015exponentially_slow_heating_floquet}.
This can be a problem, as high temperatures can interfere with the
studied effects, or even damage the systems. As described above, MFPD
operates in a regime where two major sources for heating in Floquet
engineered systems -- radiative recombination and momentum conserving
single photon absorption \citep{seetharam2015baths_controlled_floquet_population,esin2021_liquid_crystal}
-- are suppressed by the Pauli exclusion principle. Here we investigate
heating due to other relevant processes, in particular disorder and
phonon assisted single-photon absorption and interaction-assisted
single-photon absorption.

In the first process, a single electron absorbs the energy $\hbar\Omega_{F}$
and is scattered by a phonon, a defect, or an impurity. In our theory,
the rate of momentum relaxation is given by $\gamma$ (see Eq. (\ref{eq:nav_stokes})),
which captures a variety of momentum-relaxing microscopic processes.
We use $\gamma$, which is related to the quality factor $\mathcal{Q}$,
to estimate the rate of momentum non-conserving photon absorption.
In the second process, the energy of a photon is distributed between
two interacting electrons. This latter process is strongly suppressed
by the available phase-space, constrained by energy-momentum conservation.
The derivation of our estimates can be found in the Supplementary
Information.

The constant influx of energy due to these two processes heats up
the electrons. However, this influx is balanced by the cooling power
of the cold crystal lattice. A steady state is established at an effective
electron temperature $T_{e}$, where the cooling power is strong enough
to fully balance the drive-induced heating. The mechanism is dominated
by acoustic phonons and is well studied both experimentally \citep{baker2012_cooling_power_graphene_exp1,baker2013cooling_power_graphene_exp2,betz2012_cooling_power_graphene_exp_values}
and theoretically \citep{kubakaddi2009_cooling_power_graphene_theory}
for graphene. The cooling power per electron $P_{\mathrm{cool}}$
is given by 
\begin{equation}
P_{\mathrm{cool}}=\frac{\Sigma\left(\bar{\rho}\right)}{\bar{\rho}}\left(T_{e}^{4}-T_{\mathrm{ph}}^{4}\right),\label{eq:cooling_power_law}
\end{equation}
where the constant $\Sigma\left(\bar{\rho}\right)/\bar{\rho}$ plays
the role of a material dependent coupling. For a conservative estimate,
we use the values reported for graphene (see e.g. \citep{betz2012_cooling_power_graphene_exp_values}).
Theory shows that the cooling power in TMDs is even higher \citep{kaasbjerg2014_cooling_power_tmds},
such that the effective electron temperature in TMDs will be lower.
We find $\Sigma\left(\bar{\rho}\right)\approx1\frac{\mathrm{mW}}{\mathrm{K}^{4}\mathrm{m}^{2}}$
for $\bar{\rho}\approx10^{11}\mathrm{cm}^{-2}-10^{12}\mathrm{cm}^{-2}$,
$T_{\mathrm{ph}}=4.2\,\mathrm{K}$. A detailed estimation of the rates
of photon absorption (see Supplementary Information) gives a heating
power of $P_{\mathrm{drive}}=2.7\cdot10^{6}\,\mathrm{eV}/\mathrm{s}$,
for the driving strength required to reach the instability threshold
(see Discussion section). Equating $P_{\mathrm{cool}}=P_{\mathrm{drive}}$,
we find an effective electron temperature of $T_{e}\approx20\,\mathrm{K}$.
This is well below the typical temperatures in plasmonics experiments
\citep{ni2018_plasmon_quality_factor_graphene}. Moreover, due to
the nonlinear dependence of the cooling power $P_{\mathrm{cool}}$
on the electron temperature, increasing the driving power by two orders
of magnitude would result in an effective electron temperature of
only $T_{e}\approx60\,\mathrm{K}$.

All in all, we find that driving induced heating is controlled in
the setting we propose. Most importantly, we find that the driving
induced heating is too weak to significantly increase the Landau damping
of plasmons, which is strongly suppressed by the $\sqrt{q}$- shape
of the plasmon dispersion, lying -- for the frequencies considered
here -- outside the particle-hole continuum \citep{low2014graphene_plasmonics}.

\section{Discussion}

Gapped two dimensional Dirac materials are currently an active area
of research in material science \citep{chaves2020_2d_semiconductors_bandgaps,chaves2017_excitonic_tmdcs}
and promising candidates for Floquet engineering. Recently, ARPES
measurements revealed light-induced gaps in black phosphorus \citep{zhou2023black_phosphorus_floquet}
and Floquet-Bloch states in graphene \citep{merboldt2024_floquet_observation_graphene_1,choi2024_floquet_observation_graphene_2}
in the off-resonant regime considered here. Many interesting features
of MFPD, such as the enhancement of plasmon quality factors, as well
as exceptional points and non-dispersive states, appear below the
critical driving strengths determined by the instability condition
$h>2\gamma/\omega_{1}$, making them accessible for lasers with relatively
low powers. To estimate the power necessary to induce the non-equilibrium
crystalline phase, we assume typical values for the parameters of
our system: $E_{g}=0.3\,\mathrm{eV}$, $\hbar\Omega_{F}=0.35\,\mathrm{eV}$,
$\lambda=15\,\mathrm{eV}\text{Å}$. Assuming an electron density of
$\bar{\rho}=1.18\cdot10^{11}/\mathrm{cm}^{-2}$, such that the chemical
potential is close to, but above the resonance, and a plasmon quality
factor of $\mathcal{Q}=\omega_{1}/\gamma\approx10^{2}$, from the
instability condition $h>2\gamma/\omega_{1}$, we find the necessary
critical amplitude of the electric field to be $\mathcal{\bar{E}}\approx4\cdot10^{5}\,\mathrm{V}/\mathrm{m}$
with a modulation amplitude $\delta\mathcal{\bar{E}}=0.5\mathcal{\bar{E}}$.
Thus the laser intensity required to induce the transition to the
crystalline state is by a factor of $10^{4}$ smaller than in current
solid state Floquet engineering experiments \citep{mahmood2016selective_scattering_floquet-bloch_volkov,wang2013_Floquet-Bloch_states_observation,mciver2020light_anomaouls_Hall_graphene,zhou2023black_phosphorus_floquet}
and generally within the reach of continuous wave lasers. The critical
driving strength is essentially determined by the plasmon $\mathcal{Q}$.
While quality factors of $\mathcal{Q}\approx1.5\cdot10^{2}$ have
been observed in graphene \citep{ni2018_plasmon_quality_factor_graphene},
quality factors of up to $10^{3}-10^{4}$ are believed to be reachable
in principle \citep{principi2013intrinsic_graphene_plasmon_quality_factor,ni2018_plasmon_quality_factor_graphene}.
Such high quality factors would allow to further reduce the necessary
laser intensity. On the other hand, as shown in Sec. \ref{subsec:Heating-and-Landau},
the suppression of heating in the off-resonant regime used by MFPD
allows, in principle, to increase the driving power by at least two
orders of magnitude. Thus, MFPD may be even achievable with low quality
plasmons, if sufficiently strong drives are used.

The phenomena presented here can be implemented in two dimensional
Dirac systems, such as black phosphorus or TMDs \citep{manzeli2017_tmd_2d}.
Another candidate for realizing our ideas is graphene, where Floquet-Bloch
states have also been observed at different driving energies \citep{mciver2020light_anomaouls_Hall_graphene,merboldt2024_floquet_observation_graphene_1,choi2024_floquet_observation_graphene_2}.

In summary, modulated Floquet parametric driving offers a road to
realize new non-equilibrium electron phases with broken translation
symmetries in time and space and non-trivial Goldstone modes. Additionally,
MFPD opens new possibilities to excite and control THz plasmons by
optical or infrared signals. Given the interest in efficient THz technology,
the realization of plasmonic parametric amplifiers, time reversal
mirrors and other effects predicted for time-varying photonic materials
\citep{galiffi2022photonics_time_varying_ptc,lustig2018topological_photonic_time_crystal,lyubarov2022photonic_time_crystal_amplified}
will advance current research on plasmons. A direct observation of
exceptional points and non-dispersive regions can be achieved by mapping
the plasmon dispersion, e.g., with spectroscopic measurements \citep{shin2011_measuring_plasmon_dispersion_control_of_plasmon_by_doping},
while the crystalline plasmon phase can be detected via scanning near-field
optical microscopy which can be used to measure the periodic density
$\delta\rho$ of the crystalline state \citep{chen2012optical_near_field_tip_plasmons_scanning_koppens,fei2012_near_field_tip_plasmons_scanning_basov}.
The plasmon lattice could also be observed through light or electron
scattering experiments, where it would act similarly to a 2D grating.
Finally, we note that similar physics could be implemented using other
types of collective modes, e.g., magnons.

\section{Methods}

\subsection{Slowly varying envelope approximation\label{subsec:Slowly-varying-envelope}}

To find the exceptional points and instabilities of plasmon waves
under MFPD (see Sec. \ref{subsec:Instabilities-and-exceptional}),
we solve Eq. (\ref{eq:param_plasmon_driving}) in the vicinity of
the resonance $\omega_{\mathrm{pl}}\left(q\right)=\omega_{1}$ with
the slowly varying envelope approximation \citep{Landau_Lifshitz_Mechanics}.
The ansatz
\begin{equation}
\delta\rho_{q}=a_{q}\left(t\right)\cos\left(\omega_{1}t\right)+b_{q}\left(t\right)\sin\left(\omega_{1}t\right),\label{eq:slowly_varying_ansatz}
\end{equation}
when used in Eq. (\ref{eq:param_plasmon_driving}), leads to the two
equations 
\begin{align}
-\frac{1}{2}\omega_{1}^{2}hb_{q}-2\dot{a}_{q}\omega_{1}-\gamma a_{q}\omega_{1}+\left(\omega_{\mathrm{pl}}^{2}\left(q\right)-\omega_{1}^{2}\right)b_{q} & =0\nonumber \\
\frac{1}{2}\omega_{1}^{2}ha_{q}+2\dot{b}_{q}\omega_{1}+\gamma b_{q}\omega_{1}+\left(\omega_{\mathrm{pl}}^{2}\left(q\right)-\omega_{1}^{2}\right)a_{q} & =0,\label{eq:a_b_lin_eom}
\end{align}
where we neglected the second derivatives of $a_{q}$ and $b_{q}$,
as they are of higher order in the small $h$ around $\omega_{\mathrm{pl}}\left(q\right)=\omega_{1}$.
Assuming $a=a\left(0\right)e^{s\left(q\right)t}$, $b=b\left(0\right)e^{s\left(q\right)t}$,
leads to Eq. (\ref{eq:s}).

\subsection{Analysis of nonlinear pattern formation\label{subsec:Analysis-of-nonlinear_pattern}}

In this section we derive the results on the plasmon crystallization
presented in Sec. \ref{subsec:Crystallization}. We first study the
strip geometry where the $y$-axis is confined to a width $l$ with
$l<2\pi/q^{*}$. In this quasi-1D geometry, we have to replace the
Fourier transform of the Coulomb potential $V\left(q\right)$ in Eq.
(\ref{eq:plasmon_dispersion}) by
\begin{equation}
V\left(q\right)\approx\int_{0}^{l}dy\int_{-\infty}^{\infty}dx\frac{e^{i\mathbf{q}\cdot\mathbf{x}}}{\left|\mathbf{x}\right|}\approx2l\left|\ln\frac{\left|q\right|l}{4}\right|.\label{eq:quasi-1d-pot}
\end{equation}
As pointed out in Sec. \ref{subsec:Crystallization}, for the analysis
of pattern formation caused by parametric instabilities, it is sufficient
to consider a projection of Eq. (\ref{eq:nav_stokes}) on the subset
of linearly unstable modes of Eq. (\ref{eq:slowly_varying_ansatz})
\citep{zakharov1975spin_wave_turbulence,muller1994_quasipatterns_faraday}.
This excludes second order nonlinearities. These give rise to frequency
doubling modes which lie outside of the considered subspace. The third
order terms, however, contribute to oscillations at the base frequency
$\omega_{1}$. Keeping in mind that $\mathbf{u}\propto\mathbf{q}^{*}$
for plasmons, we use the linearized continuity equation $\delta\dot{\rho}=-\bar{\rho}\nabla\mathbf{u}$
to approximate
\begin{align}
u_{s,i} & \approx-\partial_{i}\frac{\delta\dot{\rho}_{s}}{q^{*2}\bar{\rho}}.\label{eq:u_through_rho}
\end{align}

At this point, it is useful to simplify Eq. (\ref{eq:rho_nonlin_eq}).
Rescaling $\delta\rho_{s}=h^{1/2}\delta\tilde{\rho}_{s}$, we find
that to order $\mathcal{O}\left(h\right)$, the time dependence of
$m^{*}$ in front of the nonlinear terms can be neglected for $h\ll1$.
With the approximations described above, for the purpose of finding
the modulated pattern of the steady state, Eq. (\ref{eq:rho_nonlin_eq})
is reduced to 
\begin{align}
\partial_{t}m^{*}\left(t\right)\partial_{t}\delta\rho_{s}+\gamma\bar{m}^{*}\partial_{t}\delta\rho_{s}+\bar{m}^{*}\omega_{1}^{2}\delta\rho_{s}\nonumber \\
-\frac{\bar{m}^{*}}{q^{*4}\bar{\rho}^{2}}\partial_{i}\partial_{j}\delta\rho_{s}\left(\partial_{i}\delta\dot{\rho}_{s}\right)\left(\partial_{j}\delta\dot{\rho}_{s}\right) & =0,\label{eq:rho_eq_on_resonance}
\end{align}
where $\delta\rho_{s}$ is given by Eq. (\ref{eq:delta_rho_ansatz}).
Derivatives (except those in parenthesis) act on all functions to
their right. An evaluation of Eq. (\ref{eq:rho_eq_on_resonance})
in which modes lying outside the subspace of linearly unstable modes
are neglected leads to Eq. (\ref{eq:a_b_nonline_eom}), which is the
nonlinear generalization of the slowly varying mode approximation
of Eq. (\ref{eq:a_b_lin_eom}).

\subsection{Deriving the Goldstone modes \label{subsec:Deriving-the-Goldstone}}

Here, we derive the dispersions of the Goldstone-like modes in the
symmetry breaking steady state. Starting from Eq. (\ref{eq:sol_phaseshift}),
we write
\begin{align}
\delta\rho & =\frac{1}{2}\sqrt{\frac{\alpha}{\beta}}\left[\sin\left(\omega_{1}t-q^{*}x-\phi\left(t,x\right)\right)\right.\nonumber \\
 & \quad\left.+\sin\left(\omega_{1}t+q^{*}x+\phi\left(t,x\right)\right)\right],\label{eq:goldstone_sol_ansatz}
\end{align}
and choose the ansatz
\begin{equation}
\phi\left(t,x\right)=\mathrm{Re}\phi_{0}e^{-i\Omega t+iQ\cdot x}\label{eq:phi_ansatz}
\end{equation}
for $\phi\left(t,x\right)$, where
\begin{equation}
\frac{Q}{q^{*}}\ll1,\ \frac{\Omega}{\omega_{1}}\ll1.\label{eq:weak_spatial_dependence}
\end{equation}
We insert Eq. (\ref{eq:goldstone_sol_ansatz}) into Eq. (\ref{eq:rho_nonlin_eq})
and compare the coefficients in front of the resulting $\sin\left[\omega_{1}t\pm q^{*}x\pm\phi\left(t,x\right)\right]$
and $\cos\left[\omega_{1}t\pm q^{*}x\pm\phi\left(t,x\right)\right]$
terms in the equations. The equation governing the propagation of
the local phase shift $\phi\left(t,x\right)$ follows from the cosine
terms. To leading order in $\varphi\left(t,x\right)$, $Q/q_{s}$
and $\Omega/\omega_{1}$, we find
\begin{equation}
\frac{\partial^{2}}{\partial t^{2}}\phi=\frac{\omega_{1}^{2}/q^{*2}}{1-h/2}\left(1+\frac{5}{2}h\right)\frac{\partial^{2}}{\partial x^{2}}\phi.\label{eq:goldstone_eq}
\end{equation}
The wave equation (\ref{eq:goldstone_eq}) leads to the dispersion
of Goldstone-like phonons (\ref{eq:Goldstone_dispersion}) in Sec.
\ref{subsec:Goldstone-like-phonons}. The equation obtained from comparing
the coefficients in front of the sine terms is fulfilled identically
to leading order in $\Omega/\omega_{1}$. A more detailed version
of this derivation can be found in the Supplementary Information.
\begin{acknowledgments}
We acknowledge useful conversations with Dmitri Basov, Gaurav K. Gupta,
Amit Kanigel, and Yiming Pan. E.K. and N.L. thank the Helen Diller
Quantum Center for financial support. N.L. is grateful for funding
from the ISF Quantum Science and Technology (2074/19) and from the
Defense Advanced Research Projects Agency through the DRINQS program,
grant No. D18AC00025. M. R. is grateful to the University of Washington
College of Arts and Sciences and the Kenneth K. Young Memorial Professorship
for support.
\end{acknowledgments}

%\bibliographystyle{thesis-bib}
%\bibliography{References_2023}

\clearpage\onecolumngrid

\setcounter{page}{1}
\section*{Supplementary Information} 

\renewcommand{\thesection}{Supplementary Sec.}%

\setcounter{figure}{0}
\renewcommand{\thefigure}{S\ \arabic{figure}}%
\renewcommand{\figurename}{Supplementary Figure}

\setcounter{equation}{0}
\setcounter{subsection}{0}
\renewcommand{\theequation}{S\,\arabic{equation}}%

\subsection{Enhancing plasmonic quality factors with MFPD}

Here we show, how MFPD below the instability threshold can be used
to enhance plasmon quality factors. Adding a forcing of the form $f=\cos\left(\omega_{1}t\right)\delta\left(q_{y}\right)$
to the right hand side of Eq. (\ref{eq:param_plasmon_driving}), we
find
\begin{equation}
\partial_{t}\left(1+h\cos\left(2\omega_{1}t\right)\right)\partial_{t}\delta\rho_{q}+\gamma\partial_{t}\delta\rho_{q}+\omega_{\mathrm{pl}}^{2}\left(q\right)\delta\rho_{q}=f\cos\left(\omega_{1}t\right)\delta\left(q_{y}\right).\label{eq:param_plasmon_driving_app}
\end{equation}
For simplicity, we assumed that the forcing is uniform along the $y$-direction
and pointlike on the $x$-axis. In real space, we have $f=\cos\left(\omega_{1}t\right)\delta\left(x\right)$.
This forcing will induce plasmons at $x=0$, which will then propagate
along the $x$-direction.

Modifying the amplitude equations (\ref{eq:a_b_lin_eom}) to include
the forcing term, and setting the derivatives $\dot{a}_{q}$, $\dot{b}_{q}$
to zero (we are looking for the steady-state solutions) we find:

\begin{align*}
-\frac{1}{2}\omega_{1}^{2}hb_{q}-\gamma a_{q}\omega_{1}+\left(\omega_{\mathrm{pl}}^{2}\left(q\right)-\omega_{1}^{2}\right)b_{q} & =f\\
\frac{1}{2}\omega_{1}^{2}ha_{q}+\gamma b_{q}\omega_{1}+\left(\omega_{\mathrm{pl}}^{2}\left(q\right)-\omega_{1}^{2}\right)a_{q} & =0.
\end{align*}
Solving for $a_{q}$ we find
\begin{align*}
a_{q} & =\frac{\gamma\omega_{1}f}{\left(\frac{\omega_{1}^{2}h}{2}\right)^{2}-\left(\omega_{\mathrm{pl}}^{2}\left(q\right)-\omega_{1}^{2}\right)^{2}-\left(\gamma\omega_{1}\right)^{2}}.
\end{align*}
\begin{align*}
b_{q} & =-\frac{1}{\gamma\omega_{1}}\left[\frac{1}{2}\omega_{1}^{2}h+\left(\omega_{\mathrm{pl}}^{2}\left(q\right)-\omega_{1}^{2}\right)\right]a_{q}\sim-\frac{\omega_{1}h}{\gamma}a_{q}
\end{align*}
For brevity, we write
\[
\omega_{\mathrm{pl}}^{2}=2Aq
\]
with 
\[
A=\frac{2\pi e^{2}\bar{\rho}}{\varepsilon\bar{m}^{*}}.
\]
For the resulting plasmon wave, we find
\begin{align*}
a\left(x\right) & =\int_{-\infty}^{\infty}dq\frac{\gamma\omega_{1}fe^{iqx}}{\left(\frac{\omega_{1}^{2}h}{2}\right)^{2}-\left(2Aq-\omega_{1}^{2}\right)^{2}-\left(\gamma\omega_{1}\right)^{2}}\\
 & =-e^{iq^{*}x}\frac{\gamma e^{-q^{*}\sqrt{\gamma^{2}/\omega_{1}^{2}-h^{2}/4}\left|x\right|}}{2A\sqrt{4\gamma^{2}-h^{2}\omega_{1}}}.
\end{align*}
Setting $h=0$ for a moment, we recognize that the damping $\gamma$
results in an exponential decay of the plasmon wave along the $x$-axis
due to the factor $e^{-\frac{q^{*}}{\mathcal{Q}}\left|x\right|}$,
where $\mathcal{Q}=\omega_{1}/\gamma$ is the quality factor. MFPD
below the instability threshold $h=2\gamma/\omega_{1}$ results in
an enhanced plasmon quality factor 
\begin{align}
\mathcal{Q}\left(h\right) & =\frac{1}{\sqrt{\gamma^{2}/\omega_{1}^{2}-h^{2}/4}}\label{eq:enhanced_quality_factor_app}\\
 & =\frac{\mathcal{Q}}{\sqrt{1-h^{2}\mathcal{Q}^{2}/4}}
\end{align}
and thereby in a longer propagation length. The enhancement effect
on the plasmon amplitude is exponential, such that even a small improvement
of $\mathcal{Q}$ will result in a strongly increased amplitude of
the plasmon field far away from the source. E.g. for the parameters
$\gamma=0.1\omega_{1}$, i.e., $\mathcal{Q}=10$ and a driving $h=\gamma/\omega_{1}$,
which is half the driving amplitude required to reach the instability
threshold, from Eq. (\ref{eq:enhanced_quality_factor_app}), we expect
an effective quality factor of $\mathcal{Q}\left(h\right)\approx11.5$.
As $h\rightarrow2\gamma/\omega_{1}$, we find $\mathcal{Q}\rightarrow\infty$.
By approaching the instability threshold, the quality of plasmon resonances
can be increased to very high values.

To support our findings, we solved for the plasmon dynamics numerically.
We first simplify the equations of motion. Starting with Eq. (\ref{eq:param_plasmon_driving_app}),
and anticipating that the amplitude variation of the plasmon wave
along the $x$-direction will be slow, i.e., the width of the plasmon
wave-packet along the $q_{x}$-axis will be small, we expand $\omega_{\mathrm{pl}}^{2}\left(q_{x}\right)$
around the resonant $q^{*}$. We find $\omega_{\mathrm{pl}}\left(q^{*}+\Delta q\right)\approx2Aq_{x}^{*}+2A\Delta q_{x}$.
Next, we approximate this dispersion with the dispersion of a massive
wave equation $\omega_{\mathrm{m}}^{2}\left(q_{x}\right)=m^{2}+Bq_{x}^{2}$.
Equating $\omega_{\mathrm{m}}^{2}\left(q^{*}+\Delta q_{x}\right)=\omega_{\mathrm{pl}}^{2}\left(q^{*}+\Delta q\right)$
to first order in $\Delta q_{x}$, gives $m^{2}=Aq^{*}$, $B=A/q^{*}$.
Thus for wavepackets centered narrowly around $q^{*}$, in real space,
Eq. (\ref{eq:param_plasmon_driving_app}) can be approximated with
the massive wave equation 
\[
\partial_{t}\left[1+h\cos\left(2\omega_{1}t\right)\right]\partial_{t}\delta\rho\left(x\right)+\gamma\partial_{t}\delta\rho\left(x\right)+\left(Aq^{*}-\frac{A}{q^{*}}\partial_{x}^{2}\right)\delta\rho_{q}=f\cos\left(\omega_{1}t\right)\delta\left(x\right).
\]
Furthermore, it is easy to see by inserting the ansatz of Eq. (\ref{eq:slowly_varying_ansatz}),
that to first order in $h$, we can approximate
\begin{equation}
\partial_{t}\left(1+h\cos\left(2\omega_{1}t\right)\right)\partial_{t}\delta\rho\left(x\right)\approx\left(1-h\cos\left(2\omega_{1}t\right)\right)\partial_{t}^{2}\delta\rho\left(x\right),\label{eq:time_varying_outside_derivative_app}
\end{equation}
since both expressions lead to the same amplitude equations (\ref{eq:slowly_varying_ansatz}).
Thus, dividing by $\left(1-h\cos\left(2\omega_{1}t\right)\right)$,
to leading order in $h\ll1$, the approximate wave equation for plasmon
propagation reads
\[
\partial_{t}^{2}\delta\rho\left(x\right)+\gamma\partial_{t}\delta\rho\left(x\right)+\left(1+h\cos\left(2\omega_{1}t\right)\right)\left(Aq^{*}-\frac{A}{q^{*}}\partial_{x}^{2}\right)\delta\rho_{q}=f\cos\left(\omega_{1}t\right)\delta\left(x\right).
\]
We use the Dedalus spectral solver \citep{burns2020dedalus} to solve
this PDE. We find that our numerical results agree very well with
the prediction of Eq. (\ref{eq:enhanced_quality_factor_app}) (see
main text).

\subsection{Plasmonic pattern formation}

\subsubsection{$\epsilon$-expansion of the amplitude equations}

To derive the gradient descent dynamics of the amplitudes towards
the steady state characterized by the minima of Eq. (\ref{eq:symmetry_breaking_potential})
of the main text, we write Eqs. (\ref{eq:a_b_nonline_eom}) of the
main text in terms of transformed variables $c=a+b$ and $d=a-b$:
\begin{align}
\dot{c} & =-\frac{\gamma}{2}c-\alpha c+\frac{\beta}{2}d\left(c^{2}+d^{2}\right)\nonumber \\
\dot{d} & =-\frac{\gamma}{2}d+\alpha d-\frac{\beta}{2}c\left(c^{2}+d^{2}\right).\label{eq:ampl_eoms-1}
\end{align}
Introducing the small parameter $\epsilon=\left(h-h_{c}\right)/h_{c}$,
and writing $\alpha_{c}=\omega_{1}h_{c}/4$, we find
\begin{align}
\dot{c} & =-\alpha_{c}\left(2+\epsilon\right)c+\frac{\beta}{2}d\left(c^{2}+d^{2}\right)\nonumber \\
\dot{d} & =\alpha_{c}\epsilon d-\frac{\beta}{2}c\left(c^{2}+d^{2}\right).\label{eq:ampl_eoms-1-1}
\end{align}
The small $\epsilon$ defines a slow time scale connected to the growth
of the unstable modes. Next, we expand in terms of $\epsilon$ and
write $c=\epsilon^{1/4}c_{1}+\epsilon^{3/4}c_{2}...$, $d=\epsilon^{1/4}d_{1}+\epsilon^{3/4}d_{2}+...$
and $\partial_{t}=\partial_{t_{0}}+\epsilon\partial_{t_{1}}+...$.
At orders $\epsilon^{1/4}$ and $\epsilon^{3/4}$, we find
\begin{align*}
\frac{\partial c_{1}}{\partial t_{0}}= & -2\alpha_{c}c_{1}\\
\frac{\partial c_{2}}{\partial t_{0}} & =-2\alpha_{c}c_{2}+\frac{\beta}{2}d_{1}\left(c_{1}^{2}+d_{1}^{2}\right).
\end{align*}
These equations describe a quick (order $\epsilon^{0}$) relaxation
towards the fixed point
\begin{align}
c_{1} & =0\nonumber \\
c_{2} & =\frac{\beta}{4\alpha_{c}}d_{1}^{3}.\label{eq:quick_partial_fixed_point}
\end{align}
At order $\epsilon^{5/4}$ we find corrections to this dynamics, which
describe how the system settles into the symmetry breaking fixed point
after $\bar{a}$ has taken the value given by Eq. (\ref{eq:quick_partial_fixed_point}).
In particular, we find
\[
\frac{\partial d_{1}}{\partial t_{1}}=\alpha_{c}d_{1}-\frac{\beta}{2}c_{2}d_{1}^{2}.
\]
Since the relaxation of $c_{2}$ towards the value given by Eq. (\ref{eq:quick_partial_fixed_point})
is fast, we can write
\begin{equation}
\frac{\partial d_{1}}{\partial t_{1}}=\alpha_{c}d_{1}-\frac{\beta^{2}}{8\alpha_{c}}d_{1}^{5}.\label{eq:grad_desc_b^5}
\end{equation}
In terms of the original variables $t$ and $\bar{b}$ the above equation
reads:
\begin{equation}
\frac{\partial d}{\partial t}=\epsilon\tilde{h}_{c}\omega_{1}d-\frac{\beta^{2}}{8\alpha_{c}}d^{5}.\label{eq:grad_desc_orig_b^5}
\end{equation}
This equation can be written as
\[
\frac{\partial d}{\partial t}=-\frac{\partial V\left(d\right)}{\partial d},
\]
with the potential $V\left(d\right)$ given by Eq. (\ref{eq:symmetry_breaking_potential})
of the main text.

\subsubsection{2D Plasmonic patterns in the symmetry breaking phase}

Whereas for the strip geometry the only feasible pattern for the crystalline
phase is a standing wave, for the 2D plane a variety of patterns is
possible. We investigated the pattern formation using the approach
of Refs. \citep{muller1994_quasipatterns_faraday,chen_vinals_1999_faraday_pattern_sel},
which analyzed pattern formation for Faraday waves. For the 2D setting
we modify the ansatz of Eq. (\ref{eq:delta_rho_ansatz}) to
\begin{align}
\delta\rho_{s}\left(t,\mathbf{x}\right) & =\cos\left(\omega_{1}t\right)\sum_{i=1}^{N}a^{\left(i\right)}\left(t\right)\cos\left(\mathbf{q}^{*\left(i\right)}\cdot\mathbf{x}\right)\nonumber \\
 & \quad+\sin\left(\omega_{1}t\right)\sum_{i=1}^{N}b^{\left(i\right)}\left(t\right)\cos\left(\mathbf{q}^{*\left(i\right)}\cdot\mathbf{x}\right).\label{eq:delta_rho_2D_ansatz_app}
\end{align}
with
\begin{equation}
\left|\mathbf{q}^{*\left(i\right)}\right|=q^{*}.
\end{equation}
The nonlinearity of the system determines the ordering pattern, which
is reflected in the number of wavevectors $N$ in Eq. (\ref{eq:delta_rho_2D_ansatz_app})
and their respective angles \citep{chen_vinals_1999_faraday_pattern_sel,chen_vinals_1997_faraday_pattern_prl,muller1994_quasipatterns_faraday}.
Typically a roll ($N=1$), a square lattice ($N=2$) or a hexagonal
lattice ($N=3$) is formed, depending on the parameters of the system.
The angles between the $\mathbf{q}_{i}^{*}$ are $\pi/2$ for the
$N=2$ and $\pi/3$ for the $N=3$ case (we note that this reasonable
assumption can fail in certain cases, e.g., for non-trivial domain
boundaries).

Inserting the ansatz (\ref{eq:delta_rho_2D_ansatz_app}) into Eq.
(\ref{eq:rho_eq_on_resonance}), we derive equations of motion analogous
to those of Eq. (\ref{eq:a_b_nonline_eom}). In equilibrium the system
is rotationally symmetric, and we assume that all standing waves in
Eq. (\ref{eq:delta_rho_ansatz}) have equal amplitudes: $a^{\left(i\right)}=a$,
$b^{\left(i\right)}=b$, for all $i$. This is a common assumption
in the analysis of pattern formation due to parametric instabilities
\citep{muller1994_quasipatterns_faraday,chen_vinals_1999_faraday_pattern_sel}.
The amplitude equations read
\begin{align*}
\dot{a} & =-\alpha b-\frac{1}{2}\gamma a+\beta_{N}b\left(a^{2}+b^{2}\right)\\
\dot{b} & =-\alpha a-\frac{1}{2}\gamma b-\beta_{N}a\left(a^{2}+b^{2}\right)
\end{align*}
and are identical with Eqs. (\ref{eq:a_b_nonline_eom}), except that
the coefficient $\beta\rightarrow\beta_{N}$ now depends on the number
of modes considered. Specifically, we find 
\begin{align}
\beta_{1} & =2\omega_{1}/32\left(\bar{\rho}\right)^{2}\label{eq:beta_N=00003D1_app}\\
\beta_{N>1}= & \left(2N-2\right)\omega_{1}/32\left(\bar{\rho}\right)^{2},\label{eq:beta_N_app}
\end{align}
where the angle between neighboring $\mathbf{q}^{*\left(i\right)}$
was assumed to be $\pi/N$, giving equal angles between all $\mathbf{q}^{*\left(i\right)}$.
In the main text {[}see Eq. (\ref{eq:symmetry_breaking_potential}){]},
we found that for a slightly supercritical driving, the system settles
in the minimum of the potential function 
\[
V\left(d\right)=-\epsilon\frac{\alpha_{c}}{2}d^{2}+\frac{\beta_{N}^{2}}{48\alpha_{c}}d^{6}
\]
when the initial growth is saturated. Here, $d=a-b$. Since the value
of $V\left(\bar{b}\right)$ at the two minima scales as $\beta_{N}^{-1}$,
we conclude that the larger the number of modes $N$, the more shallow
the minima. However, from Eqs. (\ref{eq:beta_N=00003D1_app}), (\ref{eq:beta_N_app})
follows, that the minima corresponding to standing waves $\left(N=1\right)$
and square patterns $\left(N=2\right)$ have the same depth. To settle
the question which pattern will be chosen, we have to go beyond our
approximation.

Our aim is to show that symmetry breaking and pattern formation will
occur in general, and to predict the most probable patterns. To this
aim, we use numerical simulations and leave a more involved analytical
treatment for future work. We again use the Dedalus spectral solver
\citep{burns2020dedalus} and solve equations (\ref{eq:conti})-(\ref{eq:poisson_eq})
numerically, applying the approximation of Eq. (\ref{eq:time_varying_outside_derivative_app}).
This approximation is necessary, as Dedalus does not allow to combine
time varying terms and time derivatives. The 2D Coulomb operator was
implemented in its Fourier representation $V\left(q\right)=2\pi/q$.

For the simulations we used the parameters $h=0.05$, $\gamma=0.0249\omega_{1}$.
We are thus in the strongly damped regime with $\epsilon=\left(h-h_{c}\right)/h_{c}\approx0.004$.
We used a square domain with side length of $7\lambda^{*}$, where
$\lambda^{*}=2\pi/q^{*}$ is the wavelength of the unstable modes,
and applied periodic boundary conditions. A larger domain would have
significantly increased the computational time. In all 10 runs of
our simulation the plasmons arranged themselves in square patterns.
We conclude that this is the generic steady state for a clean system.
Fig. \ref{fig:sim_patterns} of the main text shows the formation
of the square pattern starting from a smooth, random charge distribution.

\begin{figure}
\centering{}\includegraphics[scale=0.5]{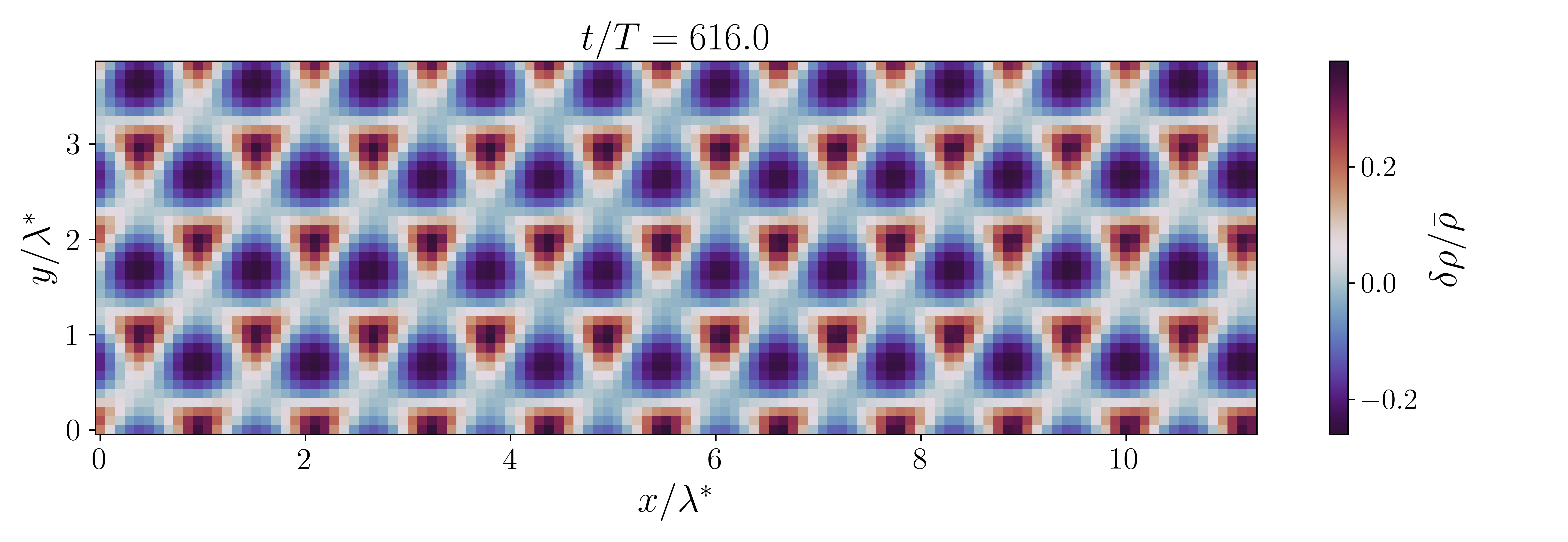}\caption{Triangular pattern observed for a sample with boundary lengths $11.32\lambda^{*}$
and $3.91\lambda^{*}$.\label{fig:Hexagonal}}
\end{figure}

Choosing incommensurate boundary length and different aspect ratios,
we found that the waves converge to a square pattern in most cases.
The pattern was mostly rotated with respect to the domain boundaries
to accommodate the periodicity of the pattern. For some aspect ratios,
the tiling was distorted, and the angle between the two modes $\mathbf{q}_{1}^{*}$,
$\mathbf{q}_{2}^{*}$ was less then 90 degrees. For a sample with
boundary length $11.32\lambda^{*}$ and $3.91\lambda^{*}$ a nearly
triangular pattern was observed (Fig. \ref{fig:Hexagonal}). However,
we could not repeat this result with other aspect ratios, and conclude
that it is the effect of fine-tuned boundary conditions.

\subsubsection{Harmonics and deviations from the square pattern}

As mentioned in the main text, the observed patterns is are not perfect
square tilings. This is due to the presence of harmonics which result
in wavemodes with wavenumbers $q=2q^{*}$ (see Fig. which shows the
Fourier spectrum of the last pattern in Fig. \ref{fig:sim_patterns}
of the main text). The origins of these modes lie beyond our analytic
treatment, however, the are clearly subleading with respect to the
main modes of the square tiling. We note that there are also very
weak contributions with wavenumbers on the order of $q\sim\sqrt{2}q^{*}$.

\begin{figure}

\begin{centering}
\includegraphics[scale=0.5]{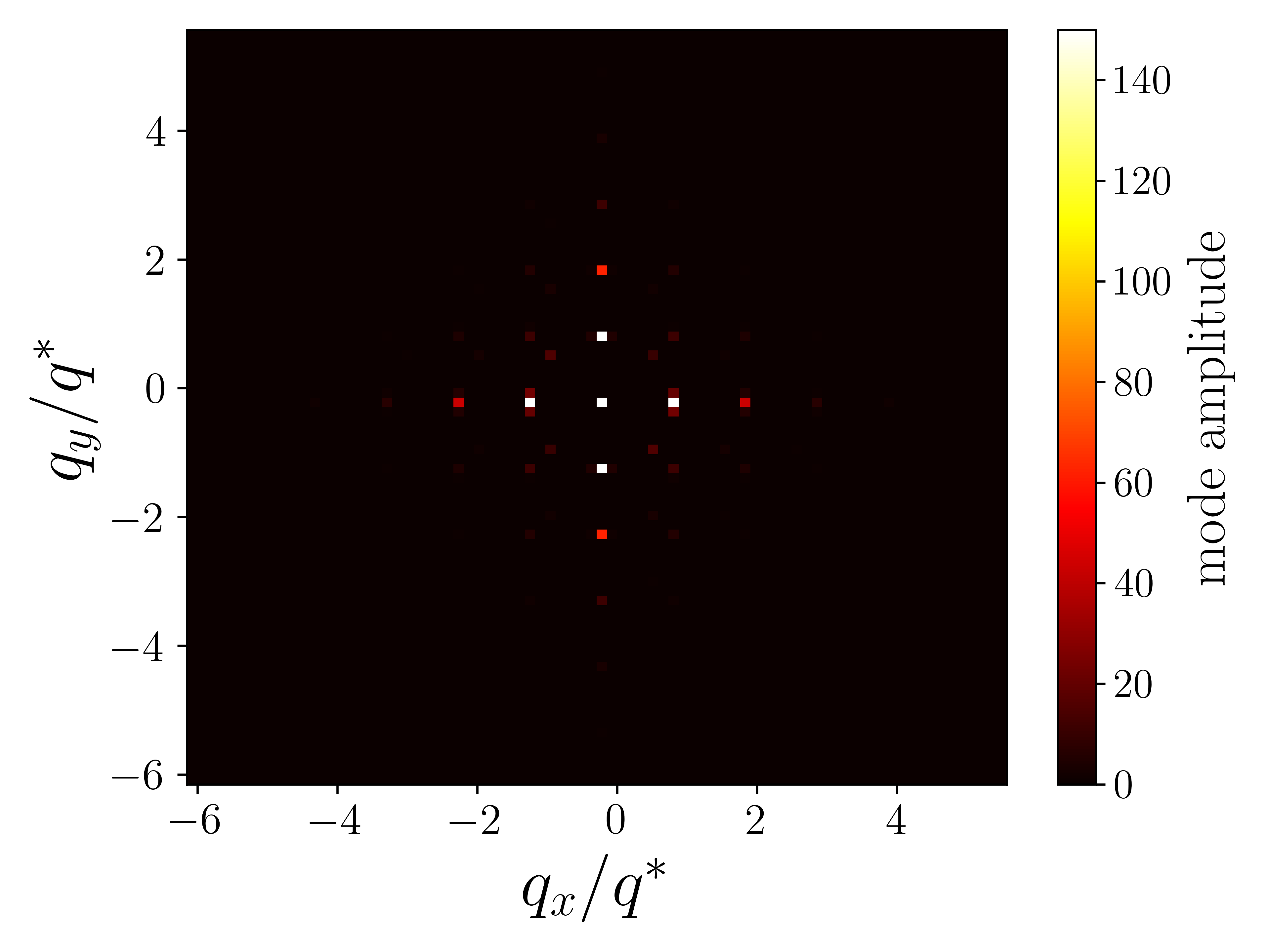}\caption{Fourier spectrum of the final pattern in Fig. \ref{fig:sim_patterns}
of the main text. While the leading modes are those corresponding
to the square tiling and wavevectors of length $q^{*}$ (closest to
the center), the spectrum is enriched by modes with wavenumbers $2q^{*}$.
\label{fig:Fourier_spectrum}}
\par\end{centering}
\end{figure}

\subsection{Goldstone-like phonons}

It is useful to rewrite the solution in Eq. (\ref{eq:sol_phaseshift})
as the sum of two moving waves
\begin{equation}
\delta\rho_{\varphi}=\frac{\sqrt{\alpha/\beta}}{2}\left[\sin\left(\omega_{1}t-q^{*}x-\varphi\left(t,x\right)\right)+\sin\left(\omega_{1}t+q^{*}x+\varphi\left(t,x\right)\right)\right].\label{eq:decomposition_right_left_waves_app}
\end{equation}
To be concise, in what follows, we will abbreviate 
\begin{align*}
\sin\left(\omega_{1}t-qx-\varphi\left(t,x\right)\right) & =\sin\left(-\right)\\
\sin\left(\omega_{1}t+qx+\varphi\left(t,x\right)\right) & =\sin\left(+\right),
\end{align*}
and similarly for the cosines. For $\varphi\left(t,x\right)$, we
choose the ansatz
\begin{equation}
\varphi\left(t,x\right)=\mathrm{Re}\varphi_{0}e^{-i\Omega t+iQ\cdot x}.\label{eq:phi_ansatz_app}
\end{equation}
A weak spatial dependence of $\varphi\left(t,x\right)$ means
\begin{equation}
\frac{Q}{q^{*}}\ll1.\label{eq:weak_spatial_dependence_app}
\end{equation}
Furthermore, we expect soft modes, such that
\begin{equation}
\frac{\Omega}{\omega_{1}}\ll1.\label{eq:slow_dynamics_of_phi_app}
\end{equation}
Eqs. (\ref{eq:weak_spatial_dependence_app}) and (\ref{eq:slow_dynamics_of_phi_app})
significantly simplify the calculations.

We first study the action of the Coulomb kernel on the Goldstone mode
$\varphi$. Linearizing $\delta\rho$ in $\varphi$ gives
\[
\delta\rho_{\varphi}\approx\sqrt{\frac{\alpha}{\beta}}\sin\left(\omega_{1}t\right)\left[\cos\left(q^{*}\cdot x\right)-\sin\left(q^{*}\cdot x\right)\varphi\left(\mathbf{x},t\right)\right].
\]
We have linearized in $\varphi$, and it is more convenient to use
the complex plain wave ansatz $\varphi\left(t,x\right)=\varphi_{0}e^{-i\Omega t+iQ\cdot x}$.
Inserting the above expression into the Coulomb integral, we find
\[
\int d^{2}x'\frac{\cos\left(\mathbf{q}^{*}\cdot\mathbf{x}\right)}{\left|\mathbf{x}-\mathbf{x}'\right|}=V\left(q^{*}\right)\cos\left(\mathbf{q}^{*}\cdot\mathbf{x}\right)
\]
and
\begin{align}
\int d^{2}x'\frac{\sin\left(\mathbf{q}^{*}\cdot x\right)\varphi\left(\mathbf{x},t\right)}{\left|\mathbf{x}-\mathbf{x}'\right|} & \approx\frac{\varphi_{0}}{2i}e^{-i\Omega t}\left(V\left(\left|\mathbf{q}^{*}+\mathbf{Q}\right|\right)e^{i\left(\mathbf{q}^{*}+\mathbf{Q}\right)\cdot\mathbf{x}}-V\left(\left|\mathbf{q}^{*}-\mathbf{Q}\right|\right)e^{-i\left(\mathbf{q}^{*}-\mathbf{Q}\right)\cdot\mathbf{x}}\right)\nonumber \\
 & \approx V\left(q^{*}\right)\sin\left(\mathbf{q}^{*}\cdot\mathbf{x}\right)\varphi\left(\mathbf{x},t\right)+\frac{1}{i}\frac{\partial V\left(q\right)}{\partial\mathbf{q}}\cdot\mathbf{\mathbf{Q}}\cos\left(\mathbf{q}^{*}\cdot\mathbf{x}\right)\varphi\left(\mathbf{x},t\right).\label{eq:transform_with_goldstone_app}
\end{align}
These expressions are valid both for the infinite 2d plane, where
\[
V\left(q\right)=\frac{2\pi}{q},
\]
and the strip geometry (with the $y$-axis confined to $0<y<l$ with
$l\ll2\pi/q^{*}$), where 
\[
V\left(q\right)=2l\left|\ln\frac{\left|q\right|l}{4}\right|
\]
and $\mathbf{q}\propto\hat{\mathbf{e}}_{x}$. We see from Eq. (\ref{eq:transform_with_goldstone_app})
that the spatial dependence of $\varphi\left(x,t\right)$ results
in a small correction of the order of $Q/q^{*}$ in the 2d plane case,
and $Q/q^{*}\ln q^{*}$ for the strip. Ignoring this correction, we
have
\[
\int d^{2}x'\frac{\delta\rho_{\varphi}}{\left|\mathbf{x}-\mathbf{x}'\right|}\approx V\left(q^{*}\right)\delta\rho_{\varphi}.
\]
Therefore, to linear order in $\phi$, we can write Eq. (\ref{eq:rho_nonlin_eq})
as {[}see Eqs. (\ref{eq:u_through_rho}) and (\ref{eq:rho_eq_on_resonance}){]}:
\begin{equation}
\partial_{t}\left(1+h\cos\left(2\omega_{1}t\right)\right)\partial_{t}\delta\rho_{\varphi}+\gamma\partial_{t}\delta\rho_{\varphi}-\frac{\omega_{1}^{2}}{q^{*2}}\partial_{i}\partial_{i}\delta\rho_{\varphi}+\frac{1}{q^{*4}\bar{\rho}^{2}}\partial_{i}\partial_{j}\delta\rho_{\varphi}\left(\partial_{i}\delta\dot{\rho}_{\varphi}\right)\left(\partial_{j}\delta\dot{\rho}_{\varphi}\right)=0.\label{eq:goldstone_gen_eq_app}
\end{equation}
We now insert Eq. (\ref{eq:decomposition_right_left_waves_app}) in
Eq. (\ref{eq:goldstone_gen_eq_app}) and neglect terms oscillating
at frequencies $3\omega_{1}$ and wavenumbers $3q^{*}$. Comparing
the coefficients in front of the $\sin\left[\omega_{1}t\pm q^{*}x\pm\varphi\left(t,x\right)\right]$
and $\cos\left[\omega_{1}t\pm q^{*}x\pm\varphi\left(t,x\right)\right]$
terms, we find the equation governing the propagation of the local
phase shift $\varphi\left(t,x\right)$. Notice that the equation obtained
by comparing the coefficients also contains corrections of order $\mathcal{O}\left(\Omega/\omega_{1}\right)$
and higher orders in $\Omega/\omega_{1}$ to the dispersion $\omega_{\mathrm{pl}}\left(q\right)$.
These describe the back action of the propagating phase $\varphi\left(t,x\right)$
on the solution $\delta\rho_{\varphi}$. Terms contributing to these
corrections can be easily recognized as they change sign depending
on the parallel or anti-parallel orientation of $Q$ with respect
to $\pm q^{*}$, i.e., for the two waves $\sin\left[\omega_{1}t\pm q^{*}x\pm\varphi\left(t,x\right)\right]$
(and similar for the $\cos$ terms). We neglect these corrections.
The equation governing the dynamics of $\varphi\left(t,x\right)$
follows from the $\cos$ terms. To leading order in $\varphi\left(t,x\right)$,
$Q/q_{s}$ and $\Omega/\omega_{1}$, we find the equation appearing
in the main text:
\begin{equation}
\frac{\partial^{2}}{\partial t^{2}}\varphi=\frac{\omega_{1}^{2}/q^{*2}}{1-h/2}\left(1+\frac{5}{2}h\right)\frac{\partial^{2}}{\partial x^{2}}\varphi.\label{eq:goldstone_eq_app}
\end{equation}

\subsection{Heating and Landau damping}

In undoped gapped systems, the main heating mechanisms are radiative
recombination and momentum conserving single photon absorption \citep{seetharam2015baths_controlled_floquet_population,esin2021_liquid_crystal}.
In the proposed set-up, these processes are suppressed by the exclusion
principle. The most relevant allowed heating processes are phonon-
or disorder-assisted, momentum non-conserving single-photon absorption
and interaction-assisted single-photon absorption. Here we estimate
the heating due to these processes and show that the Landau damping
induced by this heating is very small. 

In the following we calculate the number of electrons excited by momentum
non-conserving single-photon absorption per unit of time and area
$\Gamma_{\gamma}$, and the corresponding number for interaction-aided
single-photon absorption $\Gamma_{\mathrm{int}}$.

\subsubsection{Estimation of $\Gamma_{\gamma}$}

The matrix element for the absorption of n photons is proportional
to $\left(e\mathcal{E}\lambda/\hbar\Omega_{F}^{2}\right)^{n}$ (see,
e.g., \citep{esin2021_liquid_crystal}) and the rate of momentum non-conserving
scattering is $\gamma_{\mathrm{mom}}$. Thus, according to Fermi's
Golden rule, the scattering rate can be estimated as $\gamma_{\mathrm{mom}}\left(e\mathcal{E}\lambda/\hbar\Omega_{F}^{2}\right)^{2n}$.
To find $\Gamma_{\gamma}$, we have to multiply by the density of
electrons available for this kind of scattering. Slightly overestimating
this density (and the heating due to this process), we approximate
it by the total electron density of the upper band $\rho=\pi k_{F}^{2}/\left(2\pi\right)^{2}$.
We thus find
\begin{equation}
\Gamma_{\gamma}\approx\frac{\gamma_{\mathrm{mom}}\pi k_{F}^{2}}{\left(2\pi\right)^{2}}\left(\frac{e\mathcal{E}\lambda}{\hbar\Omega_{F}^{2}}\right)^{2}.\label{eq:rate1}
\end{equation}
Using the estimates from the main text ($\hbar\Omega_{F}=0.35\,\mathrm{eV}$,
$\lambda=15\,\mathrm{eV}\text{Å}$, $\mathcal{\bar{E}}\approx4\cdot10^{5}\,\mathrm{V}/\mathrm{m}$),
the dimensionless amplitude of the electric field controlling the
strength of the photon absorption is
\[
\frac{e\mathcal{E}\lambda}{\hbar\Omega_{F}^{2}}\approx5\cdot10^{-3}.
\]
The specific value of $\gamma_{\mathrm{mom}}$ depends on sample properties
and is unknown. However, for a conservative estimation we can set
$\gamma_{\mathrm{mom}}=\gamma$, where $\gamma$ is the decay rate
of the plasmon waves introduced in the main text. $\gamma$ can be
estimated using values for the plasmon quality factor $Q$. This is
an overestimate of $\gamma_{\mathrm{mom}}$, and hence the heating
rate, since the plasmon quality is lowered by the decay of the plasmon-induced
electric fields in the substrate -- a process which does not contribute
to $\gamma$ \citep{ni2018_plasmon_quality_factor_graphene}. A quality
factor of $Q=\omega/\gamma=10^{2}$ \citep{ni2018_plasmon_quality_factor_graphene}
gives $\gamma\approx6.28\cdot10^{10}\mathrm{Hz}$ for a plasmon frequency
of $\omega/2\pi=1\,\mathrm{THz}$. We thus find
\begin{align*}
\Gamma_{\gamma} & \approx2.5\cdot10^{-5}\cdot\gamma\rho\\
 & \approx1.6\cdot10^{6}\mathrm{Hz}\cdot\rho\cdot
\end{align*}

\subsubsection{Estimation of $\Gamma_{\mathrm{int}}$}

In this process, a single photon is absorbed and its energy is distributed
between two electrons. The two electrons exchange a momentum $\mathbf{q}$
through Coulomb interaction. Energy conservation yields
\[
\varepsilon\left(\mathbf{k}_{1}+\mathbf{q}\right)+\varepsilon\left(\mathbf{k}_{2}-\mathbf{q}\right)-\varepsilon\left(\mathbf{k}_{2}\right)-\varepsilon\left(\mathbf{k}_{1}\right)-\hbar\Omega_{F}=0.
\]
Thus, the available phase space for this process is very limited.
The Coulomb potential is given by
\[
V\left(\mathbf{q}\right)=\frac{e^{2}}{\varepsilon}\frac{2\pi}{q+q_{s}},
\]
where the screening length is $q_{s}\approx4\alpha k_{F}$, with the
fine structure constant $\alpha=e^{2}/\varepsilon\lambda$ \citep{pertsova2018excitonic_instability}.
This approximation is valid when the Fermi energy is far enough from
the bottom of the band and the dispersion is nearly linear. For the
estimate at hand, we approximate
\begin{align*}
V\left(0\right) & \approx\frac{e^{2}}{\varepsilon}\frac{2\pi}{q_{s}}\\
\varepsilon\left(\mathbf{k}\right) & \approx\lambda k
\end{align*}
The number of absorbed photons per unit of time can then be estimated
using Fermi's Golden rule as
\[
\Gamma_{\mathrm{int}}\approx\frac{1}{\hbar}\int\frac{d^{2}k_{1}}{\left(2\pi\right)^{2}}\int\frac{d^{2}k_{2}}{\left(2\pi\right)^{2}}\int\frac{d^{2}q}{\left(2\pi\right)^{2}}\delta\left(\varepsilon\left(\mathbf{k}_{1}+\mathbf{q}\right)+\varepsilon\left(\mathbf{k}_{2}-\mathbf{q}\right)-\varepsilon\left(\mathbf{k}_{2}\right)-\varepsilon\left(\mathbf{k}_{1}\right)-\hbar\Omega_{F}\right)V^{2}\left(0\right)\left(\frac{e\mathcal{E}\lambda}{\hbar\Omega_{F}^{2}}\right)^{2}.
\]
To take the integrals, it is useful to introduce dimensionless variables
\begin{align*}
\mathbf{K}_{i} & =\mathbf{k}_{i}/k_{F}\\
\mathbf{Q}' & =\mathbf{q}/k_{F}\\
\bar{\Omega} & =\hbar\Omega_{F}/\lambda k_{F}
\end{align*}
($Q'$ is not to be confused with the wavenumber of the Goldstone-like
modes, $Q$). The above integral can then be written as 
\[
\Gamma_{\mathrm{int}}=\frac{V^{2}\left(0\right)k_{F}^{6}}{\lambda\hbar k_{F}}\left(\frac{e\mathcal{E}\lambda}{\hbar\Omega_{F}^{2}}\right)^{2}\int\frac{d^{2}K_{1}}{\left(2\pi\right)^{2}}\int\frac{d^{2}K_{2}}{\left(2\pi\right)^{2}}\int\frac{d^{2}Q'}{\left(2\pi\right)^{2}}\delta\left(\left|\mathbf{K}_{1}+\mathbf{Q}'\right|+\left|\mathbf{K}_{2}-\mathbf{Q}'\right|-\left|\mathbf{K}_{2}\right|-\left|\mathbf{K}_{1}\right|-\bar{\Omega}\right).
\]
The $q$ integration can be performed by going to elliptical coordinates.
This procedure (for details see e.g. \citep{sachdev1998_elliptic_coordinates})
yields
\[
\Gamma_{\mathrm{int}}=\frac{V^{2}\left(0\right)k_{F}^{5}}{\hbar\lambda}\left(\frac{e\mathcal{E}\lambda}{\hbar\Omega_{F}^{2}}\right)^{2}\int\frac{d^{2}K_{1}}{\left(2\pi\right)^{2}}\int\frac{d^{2}K_{2}}{\left(2\pi\right)^{2}}\int\frac{d\vartheta}{\left(2\pi\right)^{2}}\left|\mathbf{K}_{1}+\mathbf{K}_{2}\right|\frac{\left(\frac{K_{1}+K_{2}+\bar{\Omega}}{\left|\mathbf{K}_{1}+\mathbf{K}_{2}\right|}\right)^{2}-\cos^{2}\vartheta}{4\sqrt{\left(\frac{K_{1}+K_{2}+\bar{\Omega}}{\left|\mathbf{K}_{1}+\mathbf{K}_{2}\right|}\right)^{2}-1}}.
\]
For the parameters of our estimate ($\hbar\Omega=0.35\,\mathrm{eV}$,
$\lambda=15\,\mathrm{eV}\text{Å}$, $\mathcal{\bar{E}}\approx4\cdot10^{5}\,\mathrm{V}/\mathrm{m}$,
$k_{F}=1.2\cdot10^{8}\,\mathrm{m^{-1}}$), we find
\[
\bar{\Omega}=1.9.
\]
A numerical evaluation of the remaining integral gives
\begin{align*}
\Gamma_{\mathrm{int}} & \approx1.02\cdot10^{-3}\cdot\frac{V^{2}\left(0\right)k_{F}^{5}}{\hbar\lambda}\left(\frac{e\mathcal{E}\lambda}{\hbar\Omega_{F}^{2}}\right)^{2}\\
 & =1.02\cdot10^{-3}\cdot\frac{V^{2}\left(0\right)k_{F}^{5}}{\hbar\lambda}\left(\frac{e\mathcal{E}\lambda}{\hbar\Omega_{F}^{2}}\right)^{2}
\end{align*}
On the other hand
\[
V\left(0\right)\approx\frac{e^{2}\pi}{2\varepsilon\alpha k_{F}}.
\]
Collecting everything and assuming $\varepsilon=6\varepsilon_{0}$,
we find
\begin{align*}
\Gamma_{\mathrm{int}} & \approx1.02\cdot10^{-3}\cdot\frac{\pi e^{4}k_{F}^{3}}{16\hbar\varepsilon^{2}\alpha^{2}\lambda}\left(\frac{e\mathcal{E}\lambda}{\hbar\Omega_{F}^{2}}\right)^{2}\\
 & \approx0.49\cdot10^{6}\,\mathrm{Hz}\cdot\rho,
\end{align*}
where we used our estimate
\[
\frac{e\mathcal{E}\lambda}{\hbar\Omega_{F}^{2}}\approx5\cdot10^{-3}.
\]

\subsubsection{Temperature estimate}

We estimate the effective temperature of electrons using experimental
data on the cooling power provided by phonons. In a Floquet driven
system, the effective temperature of the electrons and the temperature
of the lattice, which is coupled to a cold bath, can differ. The cooling
power is the amount of energy per unit of time that is carried away
by the cold lattice (mostly by optical phonons), thus stabilizing
the effective electron temperature. While experimental results for
the cooling power are available for graphene \citep{baker2013cooling_power_graphene_exp2,baker2012_cooling_power_graphene_exp1,betz2012_cooling_power_graphene_exp_values},
theoretical work \citep{kaasbjerg2014_cooling_power_tmds} shows that
the cooling power of TMDs is even larger, leading to a lower effective
temperature. Here, for a conservative estimate, we use the values
measured for graphene \citep{betz2012_cooling_power_graphene_exp_values},
where the lattice was kept at $4.2\,\mathrm{K}$.

The cooling power per electron $P$ strongly depends on the effective
temperature of the electrons. In two dimensional materials \citep{kubakaddi2009_cooling_power_graphene_theory,kaasbjerg2014_cooling_power_tmds}
one finds
\[
P_{\mathrm{cool}}=\frac{\Sigma\left(\bar{\rho}\right)}{\bar{\rho}}\left(T_{e}^{4}-T_{\mathrm{ph}}^{4}\right).
\]
This law holds well for electron temperatures below $100\,\mathrm{K}$.
$T_{e}$ is the effective electron temperature, and $T_{\mathrm{ph}}$
is the temperature of the lattice. The anove assumes that the electrons
are thermalized. In fact, one expects a very fast thermalization of
the electrons due to electron-electron interactions. For the above
estimate of $\bar{\rho}=1.18\cdot10^{11}/\mathrm{cm}^{-2}$, we find
$\Sigma\left(\bar{\rho}\right)\approx1\frac{\mathrm{mW}}{\mathrm{K}^{4}\mathrm{m}^{2}}$.
On the other hand, the power supplied by Floquet drive is 
\begin{align*}
P_{\mathrm{drive}} & =\hbar\Omega_{F}\left(\Gamma_{\mathrm{int}}+\Gamma_{\gamma}\right)/\bar{\rho}.
\end{align*}
Setting $P_{\mathrm{cool}}=P_{\mathrm{drive}}$, we find that cooling
and heating rates are balanced at the effective electron temperature
\begin{equation}
T_{e}\approx20\,\mathrm{K}.\label{eq:Temperature_estimate}
\end{equation}

Eq. (\ref{eq:Temperature_estimate}) shows that the heating induced
by MFPD is not too large. In fact, due to the nonlinear dependence
of the cooling power on $T_{e}$, the driving power could be easily
increased by two orders of magnitude giving an effective temperature
of only $T_{e}\approx60\,\mathrm{K}.$

\subsubsection{Landau damping}

The Landau damping of plasmons is described by the imaginary part
of the Lindhard function
\[
\chi\left(\omega,\mathbf{q}\right)=\int\frac{d^{2}k}{\left(2\pi\right)^{2}}\frac{f\left(\varepsilon_{\mathbf{k}-\mathbf{q}/2}\right)-f\left(\varepsilon_{\mathbf{k}+\mathbf{q}/2}\right)}{\omega+i0^{+}+\varepsilon_{\mathbf{k}-\mathbf{q}/2}-\varepsilon_{\mathbf{k}+\mathbf{q}/2}}.
\]
The imaginary part is
\[
\mathrm{Im}\left[\chi\left(\omega,\mathbf{q}\right)\right]=-\pi\int\frac{kdk}{\left(2\pi\right)^{2}}\int d\varphi\left[f\left(\varepsilon_{\mathbf{k}}\right)-f\left(\varepsilon_{\mathbf{k}+\mathbf{q}}\right)\right]\delta\left(\omega+\varepsilon_{\mathbf{k}}-\varepsilon_{\mathbf{k}+\mathbf{q}}\right).
\]
Let $\varepsilon_{\mathbf{k}}=k^{2}/2m^{*}$. This assumption simplifies
the calculations but can be made without loss of generality, because
in the end we only need to consider the linearized dispersion in the
vicinity of the Fermi surface. Defining the quantities $\mathbf{K}=\mathbf{k}/\sqrt{2m^{*}}$,
$\mathbf{Q}'=\mathbf{q}/\sqrt{2m^{*}}$(not to be confused with the
wavevector of the Goldstone-like modes, $\mathbf{Q}$), we have
\begin{align*}
\mathrm{Im}\left[\chi\left(\omega,\mathbf{q}\right)\right] & =-\pi\int\frac{kdk}{\left(2\pi\right)^{2}}\int d\varphi\left[f\left(K^{2}\right)-f\left(K^{2}+Q'^{2}+2KQ'\cos\varphi\right)\right]\delta\left(Q'^{2}+2KQ'\cos\varphi-\omega\right)\\
 & \quad-2m^{*}\pi\int\frac{KdK}{\left(2\pi\right)^{2}}\int d\varphi\left[f\left(K^{2}\right)-f\left(K^{2}+Q'^{2}+2KQ'\cos\varphi\right)\right]\delta\left(\frac{Q'^{2}-\omega}{2KQ'}+\cos\varphi\right)
\end{align*}
We are interested in the case $\frac{Q'^{2}-\omega}{2KQ'}<0$, i.e.
$-\pi/2<\varphi<\pi/2$. We substitute
\begin{align*}
\mu & =\cos\varphi\\
d\varphi & =-\frac{d\mu}{\sin\varphi}=-\frac{d\mu}{\pm\sqrt{1-\cos^{2}\varphi}}
\end{align*}
where the plus-minus-sign indicates that we have to distinguish the
cases $\varphi\lessgtr0$. We find
\begin{align}
\mathrm{Im}\left[\chi\left(\omega,\mathbf{Q}'\right)\right] & =-\frac{m^{*}\pi}{Q'}\int\frac{dK}{\left(2\pi\right)^{2}}\int_{-1}^{1}\frac{d\mu}{\sqrt{1-\mu^{2}}}\left[f\left(K^{2}\right)-f\left(K^{2}+Q'^{2}+2KQ'\mu\right)\right]\delta\left(\frac{Q'^{2}-\omega}{2KQ'}+\mu\right)\nonumber \\
 & =-\frac{m^{*}\pi}{Q'}\int_{\frac{\omega-Q^{2}}{2Q}}^{\infty}\frac{dK}{\left(2\pi\right)^{2}}\frac{1}{\sqrt{1-\left(\frac{Q'^{2}-\omega}{2KQ'}\right)^{2}}}\left[f\left(K^{2}\right)-f\left(K^{2}+\omega\right)\right],\label{eq:int_to_appr}
\end{align}
where the last step follows from 
\[
-1<\frac{Q'^{2}-\omega}{2KQ'}<0
\]
For $T=0$, approximating $f\left(K^{2}\right)-f\left(K^{2}-\omega\right)\approx-\delta\left(K-K_{F}\right)\omega/K_{F}$,
we find
\begin{align*}
\mathrm{Im}\left[\chi\left(\omega,\mathbf{q}\right)\right] & =\frac{-m^{*}\omega}{4\pi QK_{F}}\frac{\Theta\left(K_{F}-\frac{Q'^{2}+\omega}{2Q}\right)}{\sqrt{1-\frac{\left(Q'^{2}-\omega\right)^{2}}{4K_{F}^{2}Q'^{2}}}}
\end{align*}
As the theta function indicates, plasmons are undamped for
\[
\frac{qk_{F}}{m^{*}}<\omega-\frac{q^{2}}{2m}.
\]
In our case $\omega\gg q^{2}/2m^{*}$ and the condition reduces to
\[
\omega>v_{F}q.
\]

At finite temperatures and with $\omega/Q'\gg Q'$, the integral in
Eq. (\ref{eq:int_to_appr}) can be approximated as
\begin{align*}
\mathrm{Im}\left[\chi\left(\omega,\mathbf{q}\right)\right] & \approx-\frac{m^{*}\pi}{Q'}\int_{\frac{\omega}{2Q}}^{\infty}\frac{dK}{\left(2\pi\right)^{2}}\frac{K}{\sqrt{K^{2}-\left(\frac{\omega}{2Q'}\right)^{2}}}\left[f\left(K^{2}\right)-f\left(K^{2}+\omega\right)\right]
\end{align*}
For $\omega\gg v_{F}q=K_{F}Q'/2$, we can approximate the Fermi-Dirac
distributions as exponentials
\begin{align}
\mathrm{Im}\left[\chi\left(\omega,\mathbf{q}\right)\right] & \approx-\frac{m^{*}\pi}{Q'}\int_{\frac{\omega}{2Q}}^{\infty}\frac{dK}{\left(2\pi\right)^{2}}\frac{K}{\sqrt{K^{2}-\left(\frac{\omega}{2Q}\right)^{2}}}\left[e^{-\beta K^{2}}-e^{-\beta\left(K^{2}+\omega\right)}\right]\nonumber \\
 & =-\frac{m^{*}\pi}{\left(2\pi\right)^{2}Q'}\int_{\left(\frac{\omega}{2Q}\right)^{2}}^{\infty}dE\frac{1}{\sqrt{E-\left(\frac{\omega}{2Q'}\right)^{2}}}\left[e^{-\beta E}-e^{-\beta\left(E+\omega\right)}\right]\nonumber \\
 & \approx-\frac{m^{*}\omega}{4\pi Q'}\int_{\left(\frac{\omega}{2Q}\right)^{2}}^{\infty}dE\frac{\beta e^{-\beta E}}{\sqrt{E-\left(\frac{\omega}{2Q'}\right)^{2}}}=-\frac{m^{*}\omega}{4\pi Q'}\sqrt{\beta}\int_{\beta\left(\frac{\omega}{2Q}\right)^{2}}^{\infty}dx\frac{e^{-x}}{\sqrt{x-\beta\left(\frac{\omega}{2Q'}\right)^{2}}}\nonumber \\
 & =-\frac{m^{*}\omega}{4\pi Q'}\sqrt{\beta}e^{-\beta\left(\frac{\omega}{2Q'}\right)^{2}}\int_{0}^{\infty}dx\frac{e^{-x}}{\sqrt{x}}\nonumber \\
 & =-\frac{m^{*}\omega\sqrt{\beta}}{4\sqrt{\pi}Q'}e^{-\beta\left(\frac{\omega}{2Q'}\right)^{2}}\label{eq:Imag_lindhard}
\end{align}

Let us finally estimate a numerical value for the Landau damping induced
by MFPD. To find the factor $e^{-\beta\left(\frac{\omega}{2Q}\right)^{2}}$
in Eq. (\ref{eq:Imag_lindhard}), we assume $T=300\,K$ which is much
larger than the MFPD induced temperature rise. This is an overestimate,
as the temperature as well as the Landau damping will be smaller in
a typical experiment (performed at liquid nitrogen temperatures, e.g.
). For a plasmon with wavelength $1\upmu m$ and roughly $\omega=1\,\mathrm{THz}$,
we find 
\[
\beta\left(\frac{\sqrt{m^{*}}\omega}{2q}\right)^{2}\sim10^{3},
\]
such that even at room temperature, the Landau damping is suppressed
by a factor of 
\[
\sim e^{-10^{3}}.
\]

\end{document}